\documentclass[12pt]{article}

\usepackage{amssymb}
\usepackage{amsmath}
\usepackage{graphicx}
\usepackage{fullpage}
\usepackage{subfigure}
\usepackage{psfrag}
\usepackage{epsfig}

\newcommand{\cG}{\mathcal{G}}
\newcommand{\Z}{\mathbb{Z}}
\newcommand{\R}{\mathbb{R}}
\newcommand{\N}{\mathbb{N}}
\newcommand{\V}{\mathcal{V}}
\renewcommand{\v}{\mathfrak{v}}
\newcommand{\E}{\mathcal{E}}
\newcommand{\e}{\mathfrak{e}}
\newcommand{\mP}{\mathrm{Prob}}
\newcommand{\cP}{\mathcal{P}}
\newcommand{\tpi}{\tilde{\pi}}
\newcommand{\tcP}{\widetilde{\mathcal{P}}}
\newcommand{\sech}{\mathrm{sech}}

\newcommand{\Hin}[2][]{H_{#2}^{\mathrm{in}#1}}
\newcommand{\Hout}[2][]{H_{#2}^{\mathrm{out}#1}}

\newtheorem{prop}{Proposition}
\newcommand{\proof}{\noindent {\bf Proof: }}
\newcommand{\qed}{\hfill {\bf QED}\newline \vspace{5mm}}

\graphicspath{{FIGS/}{../FIGS/}{../../figures/AP/all_figs/}}

\author{Peter Ashwin\\University of Exeter,\\Exeter EX4 4QF, UK
\and 
Claire Postlethwaite\\University of Auckland,\\Auckland, New Zealand}

\title{On designing heteroclinic networks from graphs}

\begin{document}

\maketitle

\begin{abstract}
Robust heteroclinic networks are invariant sets that can appear as attractors in symmetrically coupled or otherwise constrained dynamical systems. These networks may have a complicated structure  determined to a large extent by the constraints and dimension of the system. As these networks are of great interest as dynamical models of biological and cognitive processes, it is useful to understand how particular directed graphs can be realised as attracting robust heteroclinic networks between states in phase space. This paper presents two methods of realizing arbitrarily complex directed graphs as robust heteroclinic networks for flows generated by ODEs---we say the ODEs {\em realise} the graphs as heteroclinic networks between equilibria that represent the vertices. Suppose we have a directed graph on $n_v$ vertices with $n_e$ edges. The ``simplex realisation'' embeds the graph as an invariant set of a flow on an $(n_v-1)$-simplex. This method realises the graph as long as it is one- and two-cycle free. The ``cylinder realisation'' embeds a graph  as an invariant set of a flow on a $(n_e+1)$-dimensional space. This method realises the graph as long as it is one-cycle free. In both cases we realise the graph as an invariant set within an attractor, and discuss some illustrative examples, including the influence of noise and parameters on the dynamics. In particular we show that the resulting heteroclinic network may or may not display ``memory'' of the vertices  visited.
\end{abstract}



\section{Introduction}

Heteroclinic cycle or network attractors are an interesting nontrivial form of invariant set for nonlinear dynamics where a typical trajectory recurrently approaches a number of different unstable (saddle) equilibrium states. They have been found to appear robustly in various applications \cite{krupa_97,May_Leonard1975,hofbauer_sigmund_98,field_96} and have been investigated by several groups of researchers as a way to model and understand a number of types of collective neural dynamics. In particular, the networks seem to be able to model a diverse range of systems, from (deterministic or random) sequence generation via ``winnerless competition'' \cite{Rabinovich2001} and finite-state computation \cite{seliger_tsimring_rabinovich_2003,ashwin_borresen_04}, to aspects of neural function \cite{rabinovich_afraimovich_varona_10,bick_rabinovich_2010,komarov_osipov_suykens_09} and, for example, binocular rivalry \cite{ashwin_lavric_2010}. Close dynamical relatives are found in models of phase-resetting oscillators where ``unstable attractor networks'' are a limiting case of heteroclinic networks that can appear in non-invertible semiflows \cite{AshTim05,Broetal,neves_timme_09}.

More precisely, we consider a ordinary differential equation (ODE)
$$
\dot{x}=f(x)
$$
with smooth $f(x)$ that defines a flow on $x\in\R^d$. We say an closed invariant set $X\subset \R^d$ is a {\em heteroclinic network} for this flow if it is a union of a finite set of equilibria and connecting orbits between these equilibria, such that the invariant set is chain transitive (i.e. it is a depth one heteroclinic network in the terminology of \cite{ashwin_field_99}). The set $X$ is a {\em robust heteroclinic network} if (subject to specified constraints such as symmetries and smoothness) there is an open set of perturbations of $f$ that have a nearby heteroclinic network that is homeomorphic to the original. We say a heteroclinic network is a {\em heteroclinic attractor} if it is an attractor in some sense - this could mean, for example, asymptotic stability or Milnor attraction. In the presence of additive noise, we expect trajectories starting close to such a {\em noisy heteroclinic attractor} to remain close to the network for long periods of time.

Until now, two of the main barriers to using heteroclinic networks to model specific neural processes are firstly, that coupled systems can possess quite complicated robust heteroclinic networks, and secondly, it can be highly nontrivial to determine whether a given coupled cell system has a heteroclinic network. Even when one can find a heteroclinic network, it may have very high complexity for only a moderate number of coupled cells. For example, heteroclinic networks with the structure of ``odd graphs'' \cite{AshOroBor_2010} can be found in systems of $N=2k+1$ oscillators, and heteroclinic ratchets can be found in less symmetric coupled systems \cite{karabacak_ashwin_10}. Up to now there seems to be no way to easily design a coupled dynamical system that realises a {\em given} directed network as a robust heteroclinic attractor; the closest approaches we are aware of are \cite{field_96,dias_etal_00,PD05,aguiar_ashwin_dias_field_11}. The last of these specifically considers the design of heteroclinic networks using small numbers of coupled cells by a process of ``inflation'' of cells in a smaller network but it is not clear how to use this to design an arbitrary heteroclinic network.

\begin{figure}%
\psfrag{xi1}{$x_1$}
\psfrag{xi2}{$x_2$}
\psfrag{xi3}{$x_3$}
\psfrag{xi4}{$x_1$}
\psfrag{v1}{$v_1$}
\psfrag{v2}{$v_2$}
\psfrag{v3}{$v_3$}
\psfrag{v4}{$v_4$}
\psfrag{e1}{$e_1$}
\psfrag{e2}{$e_2$}
\psfrag{e3}{$e_3$}
\psfrag{e4}{$e_4$}
\psfrag{e5}{$e_5$}
\psfrag{y1}{$y_1$}
\psfrag{y2}{$y_2$}
\psfrag{y3}{$y_3$}
\psfrag{y4}{$y_4$}
\psfrag{y5}{$y_5$}
\psfrag{p}{$p$}
\psfrag{Simplex}{Simplex realisation (2)}
\psfrag{network}{Directed graph $\cG$}
\psfrag{Cylinder}{Cylinder realisation (3,4)}
\begin{center}
\includegraphics[width=14cm]{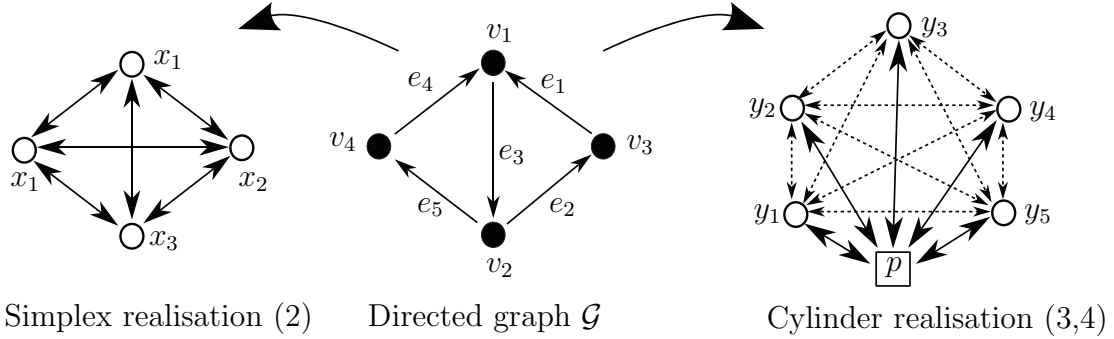}%
\end{center}
\caption{The two realisation methods described in this paper for a coupled cell network to realise a specific directed graph with $n_v$ vertices and $n_e$ edges as a heteroclinic network. The ``simplex'' realisation described in Section~\ref{sec:simplex} uses $n_v$ identical cells $x_i$ with an inhomogeneous coupling while the ``cylinder'' realisation described in Section~\ref{sec:cylinder} has a special cell ($p$) that remembers the last vertex visited while the other $n_e$ cells ($y_i$) are identical and have inhomogeneous coupling to provide dynamical connections between the states. The dashed lines indicate that the nature of coupling between the $y_i$ is of a different nature to that between that $y_i$ and $p$. The names are due to invariant structures in phase space rather than any feature of their coupling.}%
\label{fig:schem_design}%
\end{figure}

The main contribution of this paper is to give two explicit methods for design of coupled cell systems that realise a given directed graph as a heteroclinic network. In doing so, we suggest possible robust ways to embed a finite-state discrete state computational system into a system of autonomous coupled dynamical systems. The method in Section~\ref{sec:simplex} (which we call the simplex realisation) realises any graph (that is one- and two-cycle free) as a heteroclinic network for a cubic polynomial vector field on a simplex. The method in Section~\ref{sec:cylinder} (which we call the cylinder realisation) realises any graph that is one-cycle free based around a vector field on a line coupled to a number of ``transition modes''. We do this by constructing a ``coupled cell system'' that in both cases is ``all-to-all'' coupled with a small number of cell types, but where the coupling is inhomogeneous. Figure~\ref{fig:schem_design} illustrates the realisation methods.

We analyse both of these realization methods and illustrate their application to some example graphs in Section~\ref{sec:example}. In all but the simplest cases the embedded network will be part of a larger chain-recurrent set that includes additional ``induced vertices'' and separatrices between different connection sets; nonetheless we find in numerical simulations that these induced vertices do not seem to influence the the behaviour of the network, although we have no proof of this. In Section~\ref{sec:stats} we consider long-term statistics of the behaviour near the network under the influence of noise fluctuations. We see that the transition behaviour between vertices is in certain cases well-modelled as a Markov process with transition depending only on the current vertex. The presence of ``lift-off'' \cite{ArmStoKir03} of trajectories can lead to short-term memory---i.e.\ cases where the transition probability depends not only on the current vertex but also on the past few vertices visited; we lose the Markov property. This is observed in numerical simulations and discussed in Section~\ref{sec:statsnumerics}. Finally, Section~\ref{sec:discuss} finishes with a discussion of issues around the proposed methods of realization of heteroclinic cycles, and we give general thoughts on how these constructions may be related to applications.

\section{Design of heteroclinic networks realising a given directed graph}
\label{sec:design}

Suppose that $\cG=(\V,\E)$ is a directed graph between a finite set of vertices $\V=\{\v_i\}_{i=1}^{n_v}$ with directed edges $\E=\{\e_i\}_{i=1}^{n_e}$. Let $\alpha(\e_i)$ denote the starting vertex and $\omega(\e_i)$ the finishing vertex of an edge $\e_i$. We will write $A_{jk}$ to be the adjacency matrix for the graph, i.e.
$$
A_{jk} = \left\{\begin{array}{ll} 1 & \mbox{ if there is an $\ell$ with }\v_j=\alpha(\e_{\ell})
,~\v_k=\omega(\e_{\ell})\\
0 & \mbox{ otherwise.}
\end{array}\right.
$$
Now consider the noise-forced vector field on $\R^{nd}$ describing a set of coupled cells
\begin{equation}\label{eq_ode}
\frac{dx_i}{dt}=f_i(x_1,\ldots,x_n)+\zeta w_i(t)
\end{equation}
where $x_i \in \R^d$ for $i=1,\ldots,n$ and $f_i$ are smooth functions of their arguments. The terms $w_i$ represent additive i.i.d.\ noise processes and $\zeta\geq 0$ the noise amplitude. We analyse this system by considering the noise-free ($\zeta=0$) dynamics and inferring from this the low noise behaviour $\zeta\ll 1$, as well as through numerical simulations for $0<\zeta$. We say the flow induced by (\ref{eq_ode}) {\em realises} the graph $\cG$ as a heteroclinic network $X$ if $X$ is an embedding of $\cG$ such that each vertex $\v_i$ is mapped onto an equilibrium for the flow (contained in $X$), and each edge $\e_i$ is mapped onto (possibly one of a set of) heteroclinic connections from $\alpha(\e_i)$ to $\omega(\e_i)$, also contained in $X$. 

It is well known that a realisation may be robust if the form of $f$ is constrained---see for example \cite{field_96,krupa_97,golubitsky_stewart_2003,ashwin_karabacak_nowotny_2011}. The network may also be attracting if certain eigenvalue conditions are satisfied. Note that (a) robustness implies necessarily that all equilibria will be hyperbolic and (b) both of our constructions give robust realisations in that there is a a symmetry group generated by order two elements, and the constructions are robust to perturbations that respect this symmetry group.

In the remainder of this section, we present two methods for creating heteroclinic networks in a coupled cell network with $d=1$ dimensional dynamics within each cell; the coupling structures are sketched in Figure~\ref{fig:schem_design}. The first method is a cubic vector field similar to that of \cite{dias_etal_00,field_96} where the vertices of the graph are embedded as vertices of an $(n_v-1)$-simplex (note that an $n$-simplex has $n+1$ vertices)---we call this the ``simplex realisation''. Many examples of heteroclinic cycles and networks in the literature are of this type, especially the ``winnerless competition/Lotka--Volterra'' type models. Our first construction is also related to the edge and face cycle constructions given in \cite[Chapter~7]{field_96}. Other examples related to this first method are in \cite{dias_etal_00,GH88,KS94,PD05}. 

The second method, inspired by \cite[Figure 11]{ashwin_lavric_2010} has all vertices embedded on an invariant line on the centreline of an $n_e+1$ dimensional cylinder---we call this the ``cylinder realisation''. This method is also similar to the construction method in \cite{aguiar_ashwin_dias_field_11}. We will be interested in the long time (asymptotic) behaviour of the system in the presence of noise and inhomogeneities, where the heteroclinic network will not be an exact solution but it gives a clear framework in which one can discuss the dynamics of nearby solutions.

\subsection{Simplex realisation}
\label{sec:simplex}

The first construction proceeds as follows. Suppose that $\cG$ is a directed graph with $n_v$ vertices and consider the stochastically forced vector field on $x\in\R^{n_v}$ for the {\em simplex realisation} as follows:
\begin{equation}\frac{dx_j}{dt}= x_j\left(1-|x|^2+\sum_{i=1}^n a_{ij}x_i^2\right)+\zeta w_j(t)
\label{eq:simplex}
\end{equation}
with $x_j\in \R$ for $j=1,\dots, n_v$ and $|x|^2=\sum_{j} x_j^2$. The $w_j$ represent i.i.d.\ white noise modulated by $0\leq\zeta$.  In this section, we discuss the existence of the heteroclinic network $X$ in the noise free case $\zeta=0$. Note that in the absence of noise ($\zeta=0$), the system has $\Z_2^{n_v}$ symmetry, where each of the $\Z_2$ subgroups is generated by a reflection in the $x_j$th coordinate. This symmetry implies that each coordinate hyperplane is invariant under the flow (for compactness, we use the convention that a set is said to be invariant if it is invariant for the noise-free system). Note that systems such as (\ref{eq:simplex}) are known to support a wide range of heteroclinic attractors \cite{field_96}.

We say a graph $\cG$ is {\em $n$-cycle free} if is contains no directed loops of length $n$. In the following proposition we will consider graphs where there are no edges $\e_i$ with $\alpha(\e_i)=\omega(\e_i)$ (no one-cycles) and no pairs of edges $\e_{i,j}$ with $\alpha(\e_i)=\omega(\e_j)$ and $\alpha(\e_j)=\omega(\e_i)$ (no two-cycles). Note that $\cG$ is $n$-cycle free if and only if all diagonal entries of $A^n$ are zero, where $A$ is the adjacency matrix.

\begin{prop}
\label{prop:simplexembedding}
Suppose that $\cG$ is one-cycle and two-cycle free. Then $a_{ij}$ can be chosen so that the system~\eqref{eq:simplex} for $\zeta=0$ realises the graph $\cG$ as a heteroclinic network $X$. Moreover, this realisation is robust to perturbations that respect $\Z_2^{n_v}$ symmetry given by reflection in the coordinate planes.\footnote{In particular, the robustness to perturbations implies that the construction is robust to choice of parameter values.}
\end{prop}

\begin{figure}
\psfrag{xi1}{$x_i$}
\psfrag{xi2}{$x_j$}
\begin{center}
\subfigure[]{\epsfig{file=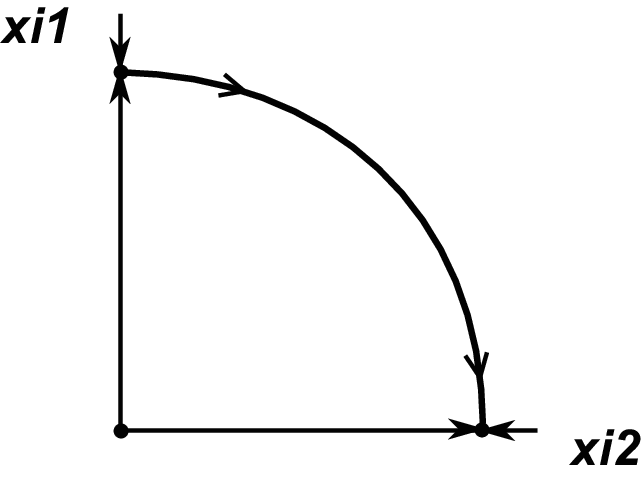,width=5cm}}
\subfigure[]{\epsfig{file=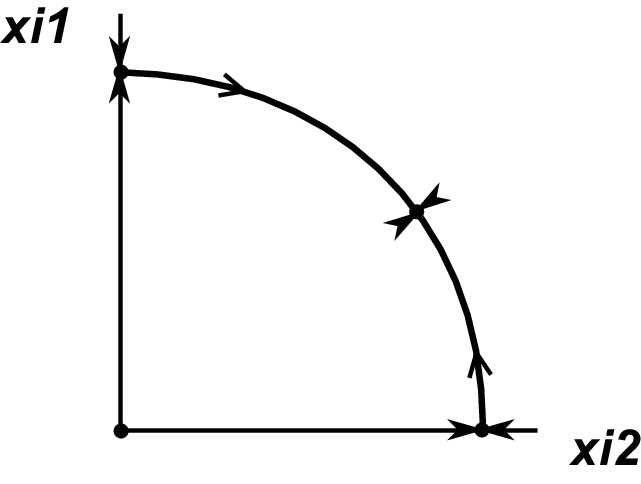,width=5cm}}
\end{center}
\caption{\label{fig:simplex_connection} (a) Schematic diagram showing a connecting orbit between equilibria $\xi_i$ and $\xi_j$ in the $(x_i,x_j)$-plane for the simplex realisation. (b) The lack of connection in this invariant subspace if $a_{ij}>0$ and $a_{ji}>0$; there is an additional equilibrium that is a sink within the subspace. Equilibria are shown with dots; the asymptotic dynamics for the full system is on topological $(n_v-1)$-simplex within an attracting invariant sphere.}
\end{figure}

\proof
The system~\eqref{eq:simplex} has equilibria on each of the $n_v$ coordinate axes,
$$
\xi_k=(0,\dots,0, 1, 0,\dots,0)
$$
where the $1$ is in the $k$th position and $k=1,\ldots,n_v$. We associate $\xi_k$ with the vertex $\v_k$ of the graph $\cG$. The eigenvalues of $\xi_k$ are $-2+3a_{kk}$ in the radial ($x_k$) direction, and $a_{kj}$ in the $x_j$ direction ($j\neq k$). We choose the coefficients $a_{ij}$ so that $a_{jj}=0$, and
$$
a_{ij}>0\mbox{ if }A_{ij}=1\mbox{ while }a_{ij}<0\mbox{ if }A_{ij}=0,\quad (i\neq j).
$$
If the graph $\cG$ contains an edge $(\e_{\ell})$ with $\alpha(\e_{\ell})=\v_i$ and $\omega(\e_{\ell})=\v_j$ (and hence does not contain an edge from $\v_j$ to $\v_i$), then $A_{ij}=1$, hence $a_{ij}>0$ and so the equilibrium $\xi_i$ will be unstable in the $x_j$ direction. Similarly, $A_{ji}=0$, thus $a_{ji}<0$ and so $\xi_j$ will be stable in the $x_i$ direction. Restricted to the two-dimensional subspace $(x_i,x_j)$, the resulting flow and connecting heteroclinic orbit are shown in Figure~\ref{fig:simplex_connection}(a). The heteroclinic network $X$ consists of the union of the equilibria $\xi_k$ and the (one-dimensional) connections from $\xi_i$ to $\xi_j$.

Because the network is one- and two-cycle free we have $A_{ij}A_{ji}=0$ for all $i,j$. If this condition were broken, for instance if $A_{ij}=A_{ji}=1$, then $\xi_i$ would be unstable to perturbations in the $x_j$ direction, and $\xi_j$ would be unstable to perturbations in the $x_j$ direction. This would result in a flow as shown in Figure~\ref{fig:simplex_connection}(b), and a new (stable) equilibrium would be created in the $(x_i,x_j)$ plane. 
\qed

This system also has a number of other equilibria; for example, the equilibrium at the origin, which is unstable (all eigenvalues are equal to $1$). The coefficients $a_{ij}$ that are non-zero can be thought of as weights on the graph; the magnitude of $a_{ij}>0$  affects the expanding eigenvalue (in the $x_j$ direction) at $\xi_i$ while $a_{ji}<0$ affects the contracting eigenvalue (in the $x_i$ direction) at $\xi_j$. In later sections we will parametrize the dependence by making the $a_{ij}$ a function of a matrix of weights $w_{ij}\geq 0$.

We conjecture that for an open set of parameter values the constructed heteroclinic network $X$ is a subset of a larger network $\tilde{X}$ that is asymptotically stable. In addition, we conjecture that in the limit of small noise $\zeta\rightarrow 0$, the only equilibria which typical trajectories will spend a non-zero proportion of time near will be the equilibria in $X$, and the only connections that will appear in non-zero frequency correspond to connections in $X$. Our numerical simulations in the later sections support these conjectures. Note that the constructed heteroclinic network $X$ from Proposition~\ref{prop:simplexembedding} cannot be an asymptotically stable heteroclinic attractor if the network is more than a cycle - this is because there will be at least one vertex with an unstable manifold that is at least two dimensional. This also means that there are typically many more connecting orbits from $\xi_i$ to $\xi_j$ than the one in $X$; this larger network $\tilde{X}$ will contain all unstable manifolds of points in the network and hence will include not only the ``edge cycles'' that represent the vertices of the graph, but also ``face cycles'' (equilibria where two or more components are non-zero) in the terminology of \cite{field_96}.

Whether this larger network $\tilde{X}$ is attracting will depend on details of the chosen parameters. Computation of the stability properties of a heteroclinic network is quite involved, see for example~\cite{KLPRS10,KPR13} or \cite{field_96} for cycles. We do not compute conditions for asymptotic stability of the embedding heteroclinic network $\tilde{X}$. However, by analogy with properties of heteroclinic cycles, we conjecture that a sufficient (but not necessary) condition for asymptotic stability of the heteroclinic network is that all the contracting eigenvalues are greater in absolute value than all the expanding eigenvalues, and all transverse and radial eigenvalues are negative. Here by contracting (resp.\ expanding) eigenvalue we mean a negative (resp.\ positive) eigenvalue in a direction associated with an incoming (resp.\ outgoing) heteroclinic connection to an equilibrium (see \cite{KM04} for cycles). This condition can presumably be achieved in this example by  choosing the $a_{ij}$ appropriately. The computation of detailed stability conditions for the network is beyond the scope of this paper, but we find in our numerical examples that the noisy heteroclinic networks can be observed to be attracting.

\subsection{Cylinder network}
\label{sec:cylinder}

The second construction also proceeds directly from the graph structure. Suppose that $\cG$ is a directed graph with with $n_e$ edges; let us define the stochastically forced vector field for the {\em cylinder realisation} on $(y,p)$ with $y\in\R^{n_e}$ and $p\in\R$ by
\begin{align}
\frac{dy_j}{dt} & = -y_j G_j(y,p)+\zeta w_j(t)
\label{eq:cylinder} \\
\frac{dp}{dt} & =-\sin(2\pi p) + F(y,p) +\zeta w_{0}(t) \nonumber
\end{align}
for $j=1,\ldots,n_e$, where $F$ and $G_j$ are smooth functions and are even in each of the $y_j$, and $F(0,\ldots,0,p)=0$ for any $p$. As before, the $w_j$ are i.i.d.\ white noise processes that we modulate with an amplitude $0\leq \zeta$. Again, we analyse the system for $\zeta=0$ while in later sections we consider numerical simulations with $0<\zeta$.

This system with $\zeta=0$ has an invariant line $y_1=y_2=\cdots=y_{n_e}=0$ parametrized by $p$, on which there are equilibria at $p=n$, $n\in\Z$, $y_j=0$. We denote the equilibrium
$$
\xi_k=(0,\ldots,0,k)
$$
for $k=1,\dots n_v$. The invariant line is contained in each of the invariant planes $\ell=1,\cdots,n_e$:
$$
P_{\ell}:=\{(y,p)~:~y_j=0~\mbox{ if }j\neq \ell\}.
$$
We say the variable $y_\ell$ is activated (i.e.\ it is non-zero and all other $y_j$ remain zero) in $P_{\ell}$, on the connection from 
$$
\alpha(\e_\ell)=(0,\ldots,0,p_{\alpha \ell})
$$
to
$$
\omega(\e_\ell)=(0,\ldots,0,p_{\omega \ell})
$$ 
where $p_{\alpha \ell}$, $p_{\omega \ell}$ are defined appropriately.
The following result holds for any graph as long as there are no edges $\e_i$ with $\alpha(\e_i)=\omega(\e_i)$---it constructs a network that lies in a cylindrical neighbourhood of the invariant line---hence the name of the network.

\begin{prop}
\label{prop:cylinderembedding}
Suppose that $\cG$ is one-cycle free. Then there are smooth functions $F$ and $G_j$ such that the system~\eqref{eq:cylinder} for $\zeta=0$ realises the graph $\cG$ as a heteroclinic network. Moreover, this realisation is robust to perturbations that respect $\Z_2^{n_e}$ symmetry given by $y_j\mapsto -y_j$.
\end{prop}

\proof
We consider the following specific choices for $F$ and $G_j$:
\begin{align}
F(y,p)&=\sum_{j=1}^{n_e} y_j^2 \tanh (p_{\omega j} -p),\nonumber \\
G_j(y,p)&=\left[ \left(y_j^2-\frac{5}{4} \right)^2 -1-f_{\alpha j}(p)  +f_{\omega j} (p)
+K_i \sum_{i\neq j} y_i^2   \right], \label{eq:cylinderaux}\\
f_{\alpha j} (p) &= L_{\alpha j} \sech^2\left(k_{\alpha j}(p-p_{\alpha j })\right), \nonumber \\
f_{\omega j} (p) & =  L_{\omega j} \sech^2\left(k_{\omega j}(p-p_{\omega j})\right).\nonumber 
\end{align}
The quantities $L_{\alpha j},  L_{\omega j}, k_{\alpha j}, k_{\omega j},K_i>0$ are parameters that will be chosen appropriately so that there is a robust heteroclinic connection in $P_{\ell}$ corresponding to the edge $\e_\ell$. Note that $K_i$ is a ``mutual inhibition'' parameter between the various $y_j$. 

 If $L_{\alpha \ell}=L_{\omega \ell}=0$ then, within the plane $P_{\ell}$, the system has $y_\ell$-nullclines at $y_\ell=0$, $y_\ell=\pm 1/2$ and $y_\ell=\pm 3/2$. The additional terms when $L_{\alpha \ell},L_{\omega \ell}\neq0$ make $\alpha(\e_\ell)$ unstable in the $y_\ell$ direction near $\alpha(\e_\ell)$. 
Constraints on the parameters can be understood by considering the geometry of the heteroclinic connection from $\alpha(\e_\ell)$ to $\omega(\e_\ell)$, which lies in $P_{\ell}$. 
\begin{itemize}
\item
Firstly, we require that $G_{\ell}>0$ (so that $\dot{y}_\ell/y_{\ell}<0$) for all $y_\ell\neq 0$ at $p\approx p_{\omega \ell}$, which is achieved if
\[
\left(y^2-\frac{5}{4} \right)^2 -1-f_{\alpha \ell}(p_{\omega \ell})+L_{\omega \ell}>0, \quad \forall y 
\]
which is satisfied if $L_{\omega \ell}>1$ and $f_{\alpha \ell}(p_{\omega \ell})$ is sufficiently small.
\item
Secondly, we require that $G_{\ell}<0$ (so that $\dot{y}_\ell/y_{\ell}<0$) when $p\approx p_{\alpha \ell }$ and $0<y_\ell<y_u$ for some $y_u>1/2$.  This is achieved if
\[
\left(y^2-\frac{5}{4} \right)^2 -1-L_{\alpha \ell}+f_{\omega \ell}(p_{\alpha \ell})<0,\quad 0<y<y_u
\]
which is satisfied if $L_{\alpha \ell}>9/16$ and $f_{\omega \ell}(p_{\alpha \ell})$ is sufficiently small.
\end{itemize}
Each of the equilibria $\xi_k$ has eigenvalues as follows: firstly, all equilibria have a `radial' eigenvalue in the $p$ direction which we label $r_k=-2\pi$.
Now, suppose  there is an edge $\e_i$ with $\alpha(\e_i)=\v_k$, then the eigenvalue at $\xi_k$ in the $y_i$ direction is
\[
e_{ki}=- \frac{9}{16}+L_{\alpha i}-f_{\omega i}(k).
\]
Similarly, if there is an edge $\e_j$ with $\omega(\e_j)=\v_k$, then the eigenvalue at $\xi_k$ in the $y_j$ direction is
\[
c_{kj} =- \frac{9}{16}+f_{\alpha j}(k)-L_{\omega j} .
\]
If the edge $\e_\ell$ neither starts nor ends at $\v_k$ (i.e.\ $\alpha(\e_\ell)\neq\v_k$ and $\omega(\e_\ell)\neq\v_k$) then the eigenvalue at $\xi_k$ in the $y_\ell$ direction will be
\[
t_{k\ell}  = -\frac{9}{16} +f_{\alpha \ell} (k)-f_{\omega \ell}(k). 
\]
Sufficient conditions for the existence of the desired connections are that $e_{ki}>0$, for all $i$ for which $\alpha(\e_i)=\v_k$ (which gives the condition $L_{\alpha i}>9/16$, as before) and $c_{kj}<0$, for all $j$ for which $\omega(\e_j)=\v_k$.

Hence we can choose parameters $L_{\alpha j}$, $L_{\omega j}$, $k_{\alpha j}$ and $k_{\omega j}$ so that $f_{\alpha j}(p)$ and $f_{\omega j}(p)$ are small enough when $p$ is far from $p_{\alpha j}$ and $p_{\omega j}$ respectively. Alternatively, we can set the values of the contracting and expanding eigenvalues (subject to some constraints; see below), and then choose parameters $L_{\alpha j}$, $L_{\omega j}$, $k_{\alpha j}$ and $k_{\omega j}$ so that the network has this set of eigenvalues. \qed

\begin{figure}%
\psfrag{x1}{$\xi_1$}
\psfrag{x2}{$\xi_2$}
\psfrag{x3}{$\xi_3$}
\psfrag{x4}{$\xi_4$}
\psfrag{y1}{$y_1$}
\psfrag{y2}{$y_2$}
\psfrag{y3}{$y_3$}
\psfrag{p}{$p$}
\begin{center}
\includegraphics[width=12cm]{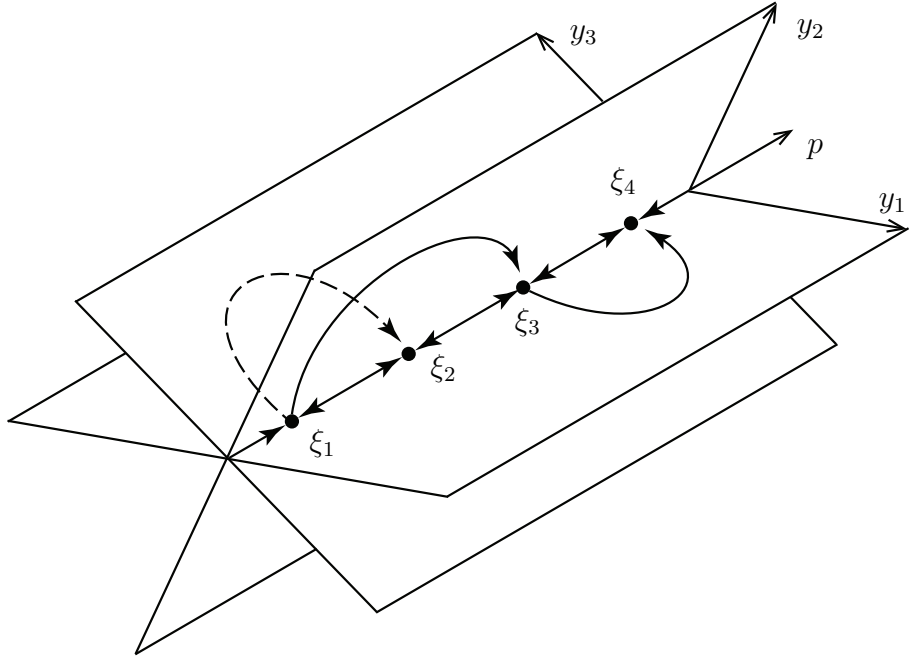}%
\end{center}
\caption{\label{fig:cylinder} 
Schematic diagram showing the structure of part of a cylinder realisation: the vertices line up along the $p$-axis while the $(y_j,p)$ planes contain connections between vertices. In this case there are connections $\xi_3\rightarrow\xi_4$ in $P_{1}$, $\xi_1\rightarrow\xi_3$ in $P_{2}$ and $\xi_1\rightarrow\xi_2$ in $P_{3}$. Note that the planes $P_{\ell}$ are all orthogonal.
}
\end{figure}

Figure~\ref{fig:cylinder} shows the structure of the connections in phase space using this construction. One implication of the requirement that $L_{\omega j}>1$ is that for these parameters, the contracting eigenvalues will always be less than $-25/16$. The final term in the $\dot{y}_j$ equations, $K_i \sum_{i\neq j} y_i^2$, is a ``mutual inhibition'' term that is included so that additional stable equilibria away from the coordinate planes are suppressed. We will typically choose $K_i=1$.

Similarly to the simplex construction, we conjecture that parameters can be chosen so that the heteroclinic network $X$ that realises $\cG$ is embedded within a larger network $\tilde{X}$; this larger network will not be an asymptotically stable attractor unless care is taken to ensure that the expanding eigenvalues are weak enough. More precisely, we conjecture that for an open set of parameters the heteroclinic network $X$ is embedded within a larger asymptotically stable network $\tilde{X}$, and in the limit of small noise $\zeta\rightarrow 0$ the only equilibria visited with non-zero proportion of time are those in $X$ and the only connections visited with non-zero frequency are those in $X$. Note that a necessary (but not sufficient) condition for asymptotic stability of the network $X$ is that $t_{k\ell}<0$.

\begin{figure}
\psfrag{p}{$p$}
\psfrag{y}{$y_\ell$}
\begin{center}
\epsfig{file=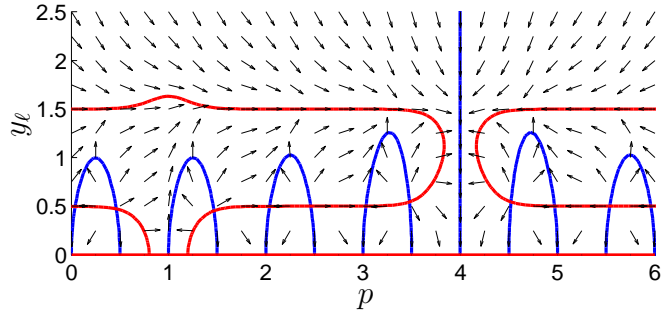,width=10cm,clip=}
\end{center}
\caption{\label{fig:cylinder_nullclines} The figure shows the nullclines and vector field for a typical two-dimensional subspace $P_{\ell}$ parametrized by $(p,y_{\ell})$ for the cylinder realisation; there is a connection from $p=1$ to $p=4$. Arrows in the vector field are all scaled to have the same length for clarity. In this case we have a connection from the saddle $\xi_1=(1,0)$ to the sink $\xi_4=(4,0)$. Note that there are additional sinks in $P_{\ell}$ at $\xi_k=(k,0)$ for $k=2,3,5$ that are surrounded by bounded basins of attraction.}
\end{figure}

Within the two-dimensional subspace $P_{\ell}$ parametrized by $(p,y_\ell)$ we show nullclines of the system with typical parameter values in Figure~\ref{fig:cylinder_nullclines}, in the case of\ $p_{\alpha \ell}=1$ and $p_{\omega \ell}=4$. The blue lines show the $p$ nullclines while the red lines show the $y_\ell$ nullclines. Note that within the subspace $y_j=0$, the equilibria $\xi_k$ ($k\neq 1$) are sinks. Also, note that there are additional equilibria at $p=k+\frac{1}{2}$, which are unstable in this one-dimensional subspace, while trajectories which start near $\xi_1$ move towards $\xi_4$.

\section{Examples of networks realised as noisy heteroclinic attractors}
\label{sec:example}

We now consider two illustrative examples; the first (the decision graph) uses the simplex realisation and Proposition~\ref{prop:simplexembedding} while the second (the Petersen graph) uses the cylinder realisation and Proposition~\ref{prop:cylinderembedding}. The decision graph could be realised using the cylinder realisation, but the Petersen graph cannot be realised using the simplex realisation as it contains cycles of order two. In both cases we consider simulations that include small but non-zero levels of noise.

\subsection{Decision graph}
\label{sec:decision}

\begin{figure}%
\psfrag{v1}{$v_1$}
\psfrag{v2}{$v_2$}
\psfrag{v3}{$v_3$}
\psfrag{v4}{$v_4$}
\psfrag{v5}{$v_5$}
\psfrag{v6}{$v_6$}
\psfrag{v7}{$v_7$}
\psfrag{v8}{$v_8$}
\centerline{\includegraphics[width=6cm]{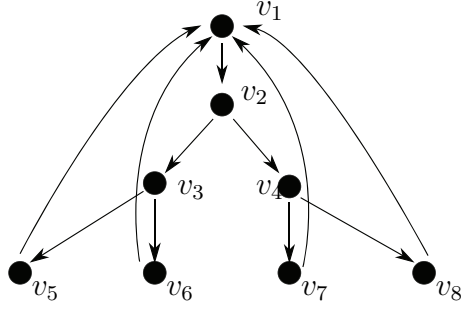}}
\caption{The ``decision network'' graph for the adjacency matrix corresponding to the weights in (\ref{eq_graph1_weights}). 
\label{fig:decisiongraph}}%
\end{figure}

Consider the ``decision tree with reset'' shown in Figure~\ref{fig:decisiongraph}, which for convenience we call the ``decision graph''. One can think of this graph $\cG$ as a two-level binary decision tree followed by a reset back to the first (root) vertex of the tree. This graph has $n_v=8$ vertices and $n_e=11$ edges. Its adjacency matrix is $A_{ij}=1$ for $w_{ij}\neq 0$ and $A_{ij}=0$ for $w_{ij}=0$, where
\begin{equation}
\{w_{ij}\}=\left[\begin{array}{cccccccc}
0 & a & 0 & 0 & 0 & 0 & 0 & 0\\
0 & 0 & a & a' & 0 & 0 & 0 & 0\\
0 & 0 & 0 & 0 & a & a' & 0 & 0\\
0 & 0 & 0 & 0 & 0 & 0 & a & a'\\
a & 0 & 0 & 0 & 0 & 0 & 0 & 0 \\
a & 0 & 0 & 0 & 0 & 0 & 0 & 0 \\
a & 0 & 0 & 0 & 0 & 0 & 0 & 0 \\
a & 0 & 0 & 0 & 0 & 0 & 0 & 0
\end{array}\right].
\label{eq_graph1_weights}
\end{equation}
If we set parameters
\begin{equation}
\label{eq:decision_params}
a=0.99,~a'=0.98,~\sigma=0.2,~\mu=0.6,~\zeta = 10^{-4}
\end{equation}
and define
$$
a_{ij}=\left\{
\begin{array}{ll}
-\sigma+\mu w_{ij} & \mbox{ if }i\neq j\\
0 & \mbox{ otherwise}
\end{array}\right.
$$
then it is a routine calculation to check that the requirements of Proposition~\ref{prop:simplexembedding} are satisfied and (\ref{eq:simplex}) will realise a noisy heteroclinic network with the structure of the graph shown in Figure~\ref{fig:decisiongraph}.

\begin{figure}%
\psfrag{x_1^2}{$x_1^2$}
\psfrag{x_2^2}{$x_2^2$}
\psfrag{x_3^2}{$x_3^2$}
\psfrag{x_4^2}{$x_4^2$}
\psfrag{x_5^2}{$x_5^2$}
\psfrag{x_6^2}{$x_6^2$}
\psfrag{x_7^2}{$x_7^2$}
\psfrag{x_8^2}{$x_8^2$}
\psfrag{t}{$t$}
\centerline{\includegraphics[width=14cm]{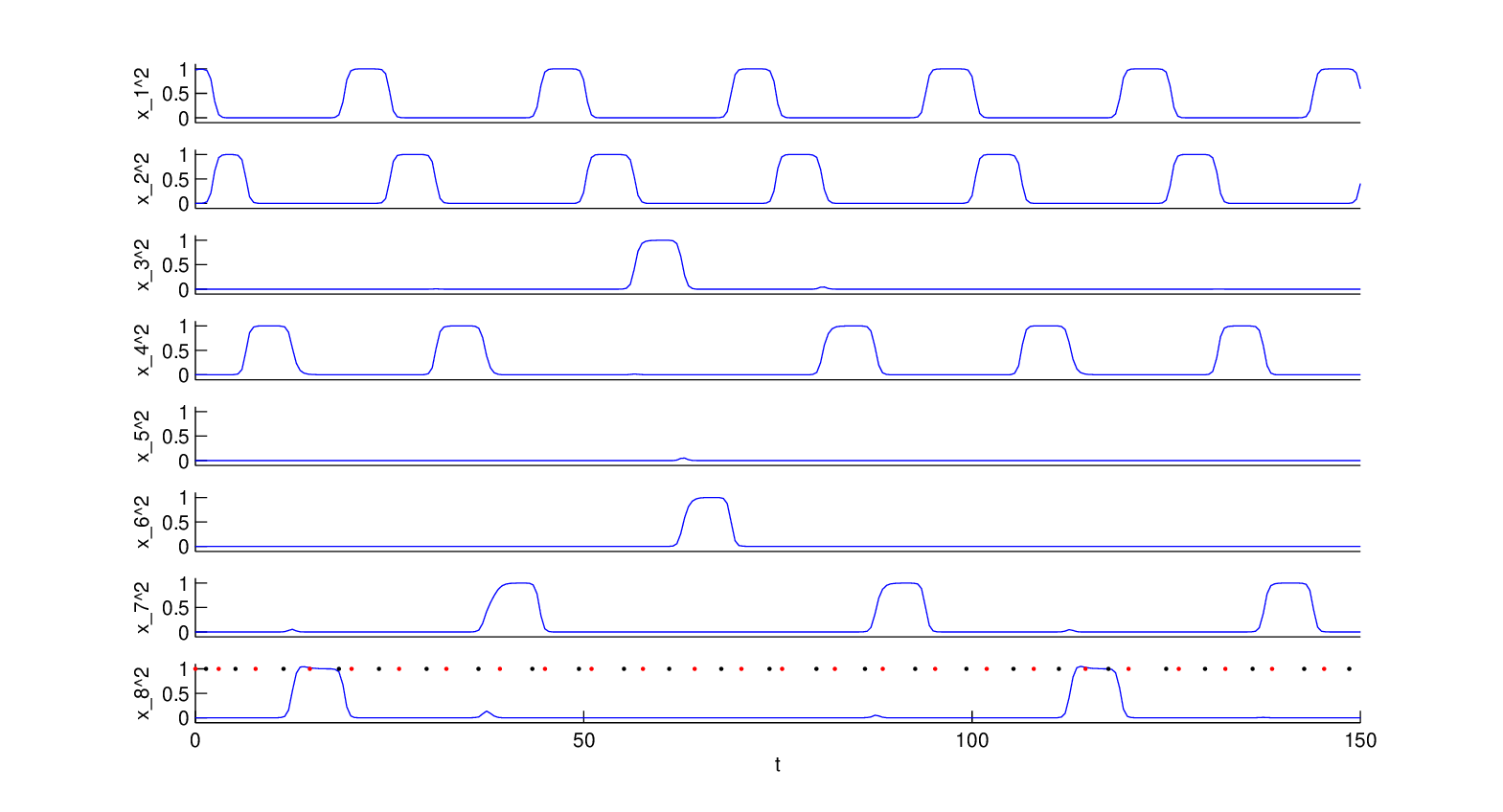}}
\psfrag{xi1}{$\xi_1$}
\psfrag{xi2}{$\xi_2$}
\psfrag{xi3}{$\xi_3$}
\psfrag{xi4}{$\xi_4$}
\psfrag{xi5}{$\xi_5$}
\psfrag{xi6}{$\xi_6$}
\psfrag{xi7}{$\xi_7$}
\psfrag{xi8}{$\xi_8$}
\psfrag{Rr}{\raisebox{-0.3cm}{$\mathrm{Re}(R)$}}
\psfrag{Ri}{\hspace{-0.3cm}$\mathrm{Im}(R)$}
\centerline{\includegraphics[width=8cm]{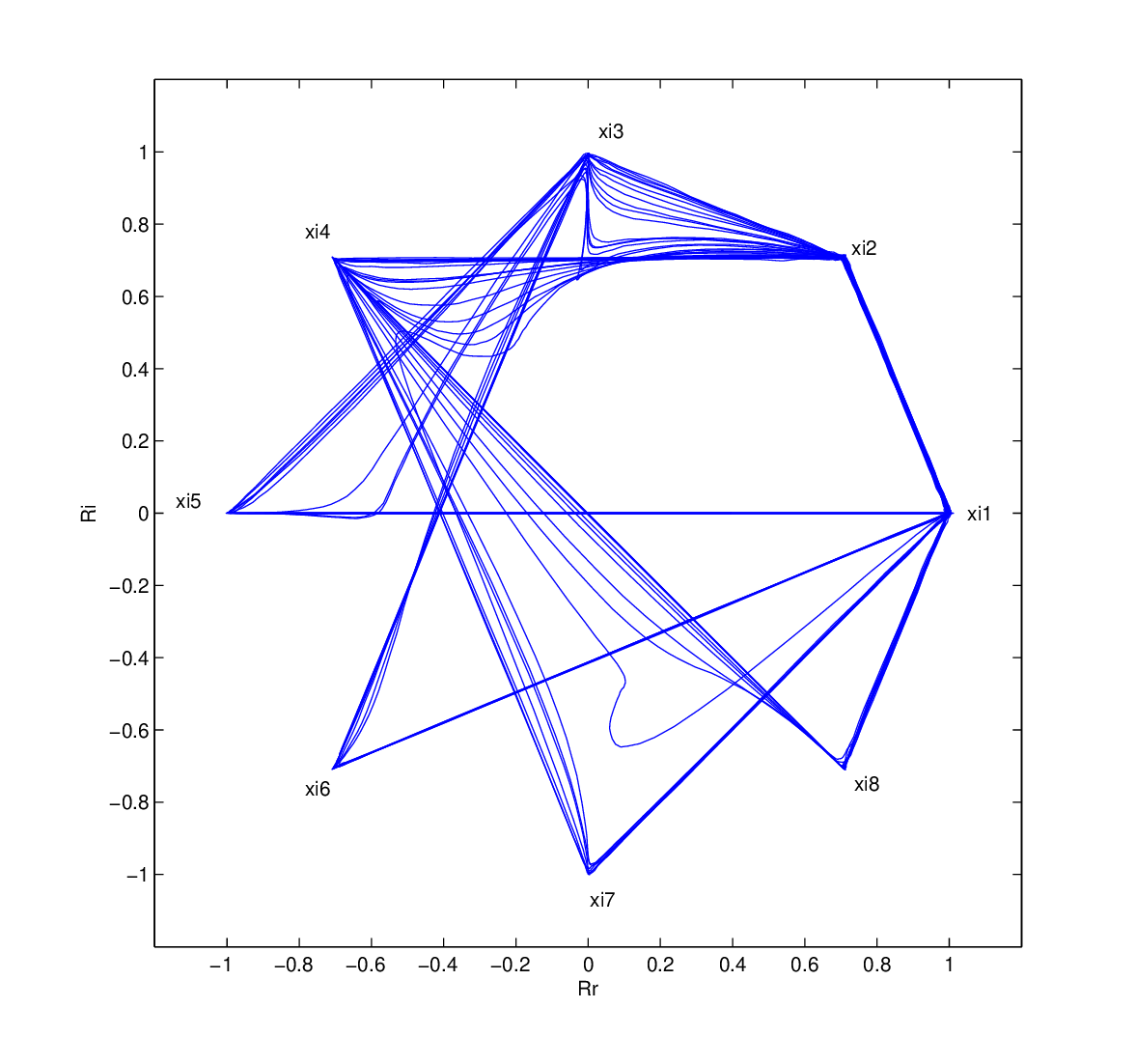}}
\caption{An example simulation of the decision graph (shown in Figure~\ref{fig:decisiongraph}) using the simplex realisation~\eqref{eq:simplex} and weights (\ref{eq_graph1_weights}). Upper: time series of $x_i^2$; the red (resp. black) dots on the lowest frame indicate times where the trajectory enters (resp. leaves) an ``epoch'' of being close to one of the saddles. Lower: projection onto the real and imaginary parts of order parameter for the system showing the location of the equilibria. Simulations clearly show a noisy heteroclinic attractor with the structure as in Figure~\ref{fig:decisiongraph}, projected into the plane using the complex observable $R(t)$ (equation~\eqref{eq:R}). Note that presence of additional saddles can be inferred from rare excursions away from the one-dimensional connections.\label{fig:decisiongraph_project} }%
\end{figure}

One can visualise the dynamics on the realisation of this graph by projecting the trajectory onto two observables that order the vertices into a ring. More precisely, for a trajectory $x(t)$ we define the complex observable
\begin{equation}\label{eq:R}
R(t) = \frac{1}{n_v}\sum_{k=1}^{n} x_k^2 \exp\left[i\pi \frac{2(k-1)}{n_v}\right]
\end{equation}
so that the vertices of the graph are projected onto the vertices of a regular $n_v$-gon on the unit circle. Figure~\ref{fig:decisiongraph_project} shows a projection and timeseries for the dynamics of the system corresponding to the graph in Figure~\ref{fig:decisiongraph}.

Proposition~\ref{prop:simplexembedding} means that the graph structure is unaffected by the exact values of the parameters used. However, the statistics of the residence times and the transition probabilities at decision points (including the possibility of memory) are strongly affected by the parameter values. We explore some of these effects in Section~\ref{sec:stats} and apply our results to investigate this graph in Section~\ref{sec:statsnumerics}.

\subsection{Petersen graph}
\label{sec:petersen}

\begin{figure}%
\psfrag{v1}{$v_1$}
\psfrag{v2}{$v_2$}
\psfrag{v3}{$v_3$}
\psfrag{v4}{$v_4$}
\psfrag{v5}{$v_5$}
\psfrag{v6}{$v_6$}
\psfrag{v7}{$v_7$}
\psfrag{v8}{$v_8$}
\psfrag{v9}{$v_9$}
\psfrag{v10}{$v_{10}$}
\centerline{\includegraphics[width=6cm]{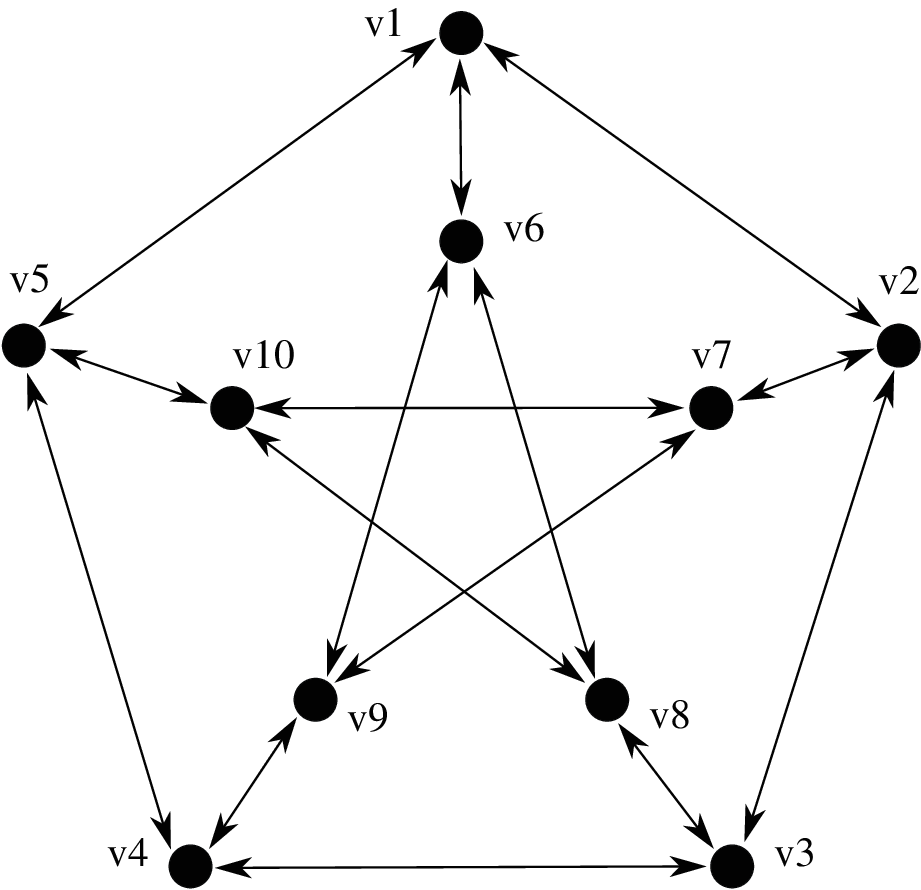}}
\caption{The Petersen graph; this can be viewed as a (one-cycle free) directed graph by considering  each edge as a pair of directed edges in both directions.}%
\label{fig:petersengraph}%
\end{figure}

As an example with a somewhat different structure (and one that is nontrivially but highly connected), we now consider a cylinder realisation of the Petersen graph (shown in Figure~\ref{fig:petersengraph}) as a noisy heteroclinic network. Each edge can be thought of as a pair of directed edges, one in each direction. 

This graph has been studied extensively as one of the simplest examples of a graph with certain nontrivial colouring properties; it has also been found to organize heteroclinic networks in five globally coupled phase oscillators and in systems of five delay pulse-coupled oscillators, where it has been proposed for computational purposes \cite{ashwin_borresen_04,neves_timme_12}.

\begin{figure}%
\psfrag{p}{$p$}
\psfrag{t}{\raisebox{-0.1cm}{$t$}}
\psfrag{y_i^2}{$y_i^2$}
\centerline{\includegraphics[width=14cm,clip=]{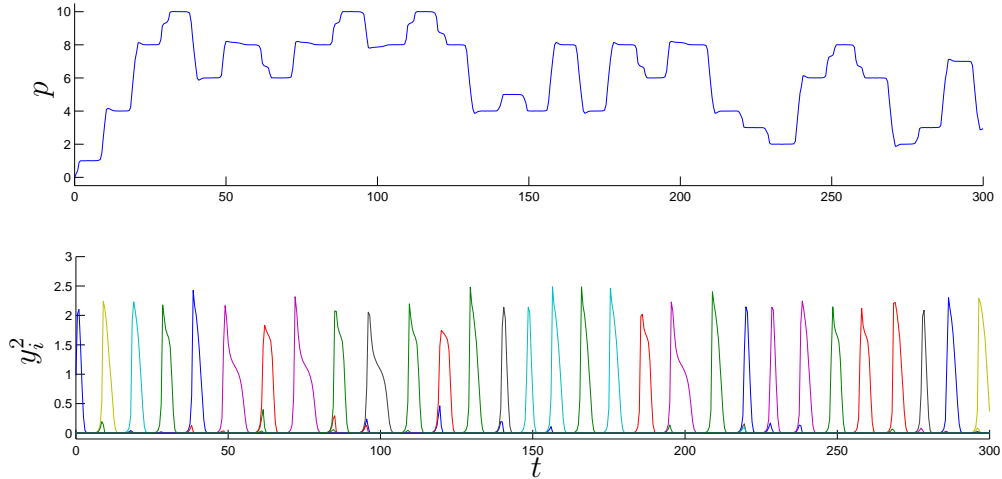}}
\caption{Time series for a cylinder realisation of the Petersen graph (Figure~\ref{fig:petersengraph}). The $p$-dynamics (top) can be observed to wander around a noisy heteroclinic attractor between the vertices of the Petersen graph, only making transitions corresponding to edges in the graph, while each of the components $y_i$ for $i=1,\ldots,15$ (bottom) become non-zero only during a transition along the $i$th edge. The presence of weak noise causes the dynamics to wander around the network. Note that all the saddles corresponding to vertices on the network have three dimensional unstable manifolds.}
\label{fig:petersents}
\end{figure}

Figure~\ref{fig:petersents} shows a simulation of a realisation of this using the cylinder realisation (\ref{eq:cylinder},\ref{eq:cylinderaux}) and adjacency matrix
\begin{equation}
\{A_{ij}\}=\left[\begin{array}{cccccccccc}
0 & 1 & 0 & 0 & 1 & 1 & 0 & 0 & 0 & 0\\
1 & 0 & 1 & 0 & 0 & 0 & 1 & 0& 0 & 0\\
0 & 1 & 0 & 1 & 0 & 0 & 0 & 1& 0 & 0\\
0 & 0 & 1 & 0 & 1 & 0 & 0 & 0& 1 & 0\\
1 & 0 & 0 & 1 & 0 & 0 & 0 & 0 & 0 & 1
\\
1 & 0 & 0 & 0 & 0 & 0 & 0 & 1 & 1 & 0\\
0 & 1 & 0 & 0 & 0 & 0 & 0 & 0 & 1 & 1\\
0 & 0 & 1 & 0 & 0 & 1 & 0 & 0 & 0 & 1\\
0 & 0 & 0 & 1 & 0 & 1 & 1 & 0 & 0 & 0\\
0 & 0 & 0 & 0 & 1 & 0 & 1 & 1 & 0 & 0
\end{array}\right].
\label{eq_petersen_weights}
\end{equation}
The other parameters used for the simulation in Figure~\ref{fig:petersents} are
\begin{equation}
\label{eq:petersen_params}
L_{\alpha j}=1.4376,~L_{\omega j}=1.5625, k_{\alpha j}=2.017,~k_{\omega j}=0.4705,~K_i=1,~\zeta=10^{-6}.
\end{equation}
Note that by varying $k_{\alpha j}$ and $k_{\omega j}$ with $j$ we can vary the expanding and contracting eigenvalues on the $j$th connection. For simplicity, in the example we set them to be identical.

\subsection{Further examples}

\begin{figure}%
\psfrag{v1}{$v_1$}
\psfrag{v2}{$v_2$}
\psfrag{v3}{$v_3$}
\psfrag{v4}{$v_4$}
\psfrag{v5}{$v_5$}
\psfrag{(a)}{(a)}
\psfrag{(b)}{(b)}
\centerline{\includegraphics[width=10cm]{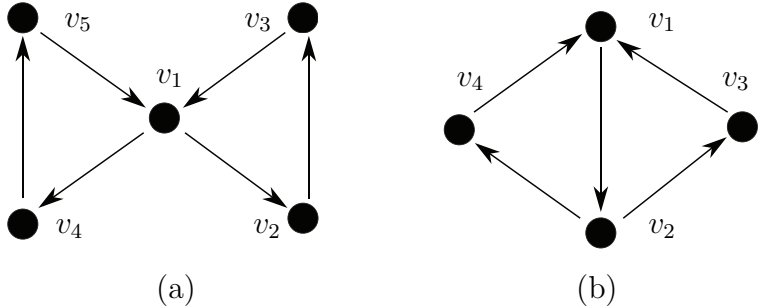}}
\caption{(a) A ``Bowtie'' graph of two connected cycles and (b) the ``Kirk--Silber''~\cite{KS94} network that shows a competition of two length three cycles with a shared edge. Note that in both cases there are two cycles of length three; in (b) there is a shared edge and two shared vertices while in (a) there is only a shared vertex.
}%
\label{fig:bowtie_KS}%
\end{figure}

We briefly highlight two further graphs that illustrate different ways in which vertices can connect cycles of the same type within a network. Figure~\ref{fig:bowtie_KS} illustrates two networks each containing two nontrivial cycles of length three. In each case the network can in principle be realised using either ``simplex'' or ``cylinder'' realisations. Note that although the decision graph has alternative routes around the graph, all of these share at least one edge; similarly the cycles on the Kirk--Silber network share an edge, while those in the Bowtie graph do not.

\section{Statistical properties of trajectories near a realised network}
\label{sec:stats}

Propositions~\ref{prop:simplexembedding} and \ref{prop:cylinderembedding} do not give unique ways to realise a given graph $\cG$ as a heteroclinic attractor - rather, they give open sets of functions that give the appropriate embedding. These alternative realisations can however display different statistical properties of trajectories near the network in the presence of low noise (or more generally, time-dependent inputs) after transients have decayed. In this section, we adapt and generalise some of the analysis of Stone, Armbruster and colleagues~\cite{StoHol90,StoArm99,ArmStoKir03} to understand some basic statistical properties of the noise-induced itineraries of trajectories near the network.

We fix on a size $h>0$ for neighbourhoods of the equilibria and say a solution trajectory $x(t)$ is {\em close} to an equilibrium $\xi_j$ (corresponding to a vertex of the realised graph) at time $\tilde{t}$ if $|x(\tilde{t})-\xi_j|<h$. We require that $|\xi_p-\xi_q|>2h$ for any $p\neq q$ so that at any time, the trajectory is close to at most one equilibrium. We say $x(t)$ {\em remains close to} $\xi_j$ during $t\in[s,s+T]$ if it satisfies
$$
|x(t)-\xi_j|<h\mbox{ for }t\in(s,s+T)\mbox{, and } |x(t)-\xi_j|=h\mbox{ for }t\in\{s,s+T\}.
$$
We divide a trajectory starting at some given initial condition $\tilde{x}$ and given noise path $\tilde{\vartheta}$ into an {\em itinerary}, i.e. the sequence of {\em epochs}:
\begin{equation}\label{eq:itinerary}
\{(i_k,s_k,T_k)~:~k\in\N\}
\end{equation}
such that $x(t)$ remains close to $\xi_{i_k}$ during $t\in[s_k,s_k+T_k]$. Note that $i_k\in\{1,\ldots,n_v\}$ represent indicies of the vertices (equilibria), the times of entry $0<s_k$ are increasing and the durations $T_k>0$ are all positive; for example, the red and black dots in Figure~\ref{fig:decisiongraph} illustrate the sequences $s_k$ and $s_k+T_k$ respectively, for $h=0.1$. 

One can formally consider the solution of the noise-perturbed system as deterministic in the spirit of random dynamical systems \cite{Arnold98}, i.e. as a skew product evolution
\begin{equation}
\begin{array}{rcl}
x(t) & = & \Phi_t(\tilde{x},\tilde{\vartheta})\\
\vartheta(t) &=& \theta_t(\tilde{\vartheta})
\end{array}
\label{eq:skewprod}
\end{equation}
over the evolution $\theta_t$ along the noise path. In the original system $\tilde{\vartheta}$ represents a particular Brownian path whereas in the simulation, $\tilde{\vartheta}$ can be thought of as choice of seed for a random number generator. Note that $\Phi_t$ satisfies the cocycle property $\Phi_{t+s}(\tilde{x},\tilde{\vartheta})=\Phi_{t}(\Phi_{s}(\tilde{x},\tilde{\vartheta}),\theta_s(\tilde{\vartheta}))$. Then $x(t)$ remains close to $\xi_j$ if $|\Phi_t(\tilde{x},\tilde{\vartheta})-\xi_j|<h$, which clearly is determined by initial condition and noise path.

If the trajectory remains close to the attractor then it will have an infinite itinerary except in the (very unlikely) case that it remains close to one of the saddle equilibria for all time. We make a {\em stochastic stationarity} assumption that the statistical properties of the itinerary of typical initial conditions and typical noise trajectory are stationary and independent of the initial condition and details of the noise trajectory. This means we assume that any transients associated with the initial condition will have decayed and that the initial condition is distributed according to a stationary invariant probability distribution on the space of initial conditions $\tilde{x}$ for the noise-perturbed system, and the noise path $\tilde{\vartheta}$ is ``typical''. More precisely, we make an implicit assumption that there is an ergodic probability measure for $\theta_t$ that lifts to a natural measure for the skew product flow. With respect to this distribution, we define the probability of observing a given finite sequence of vertices $\{j_k~:k=1,\ldots,m\}$ as
\begin{equation}\label{eq:fiprob}
\cP(j_1,\ldots,j_m)=\mP(~(\tilde{x},\tilde{\vartheta})~:~i_{\ell}=j_\ell~\mbox{ for }~\ell=1,\ldots,m \mbox{, for the trajectory starting at $(\tilde{x},\tilde{\vartheta})$}).
\end{equation}
The stationarity assumption implies that one can compute this from a typical trajectory simply in terms of the frequency of transitions
\begin{equation}\label{eq:approxprob}
\cP(j_1,\ldots,j_m)=\lim_{k\rightarrow \infty} \frac{1}{k} \#\{ 0 \leq \ell < k ~:~ i_{\ell+n}=j_n \mbox{ for }n=1,\ldots,m\}.
\end{equation}
This can be used to define the probability $\pi_j$ of an epoch being close to $\xi_j$:
$$
\pi_j= \cP(j)
$$
and (assuming $\pi_j>0$) we define the transition probability between vertices $\xi_{j_1}$ and $\xi_{j_2}$ by:
\begin{equation}\label{eq:tprob}
\pi_{j_1,j_2}=\frac{\cP(j_1,j_2)}{\pi_{j_1}}.
\end{equation}
We ask the question: are the statistics of the itineraries simply those of a Markov chain whose non-zero transitions correspond to edges of the original graph $\cG$? Although this is possible, memory effects can appear (see~\cite{Bak2010}) where transition probabilities may depend not just on the current equilibrium but also on previous equilibria visited.  

We say a noise perturbed attracting heteroclinic network is {\em memoryless} if the transition probabilities are independent of the prior itinerary, i.e. if the itineraries are Markov of order one. More precisely, a transition from $\xi_p$ to $\xi_q$ is said to be memoryless if 
\begin{equation}\label{eq:memoryless}
\cP(j_1,\ldots,j_m)= \cP(j_1,\ldots,j_{m-1})\pi_{p,q}
\end{equation}
for any $m\geq 2$ and any sequence $\{j_k~:k=1,\ldots,m\}$ with $j_{m-1}=p$ and $j_m=q$. The system is memoryless if all possible transitions are memoryless. We expect such a system will be truly memoryless only in the singular limit of low noise $\zeta\rightarrow 0$; nonetheless, later sections suggest that a system can be very close to memoryless in that corrections to (\ref{eq:memoryless}) may be asymptotically small in $\zeta$. Note also that a particular transition will necessarily be memoryless if $\pi_{p,q}=1$, though this does not necessarily imply that all other transitions are memoryless.

The transition probabilities will be affected (to a greater or lesser extent) by the eigenvalues at the equilibria and by the noise level. In the following section, we outline the influence of these parameters. Finally, using the simplex realisation of the decision  graph, 
Section~\ref{sec:statsnumerics} gives some numerical examples with and without memory.

\subsection{Analysis of dynamics near heteroclinic networks: Poincar\'e sections and maps}

Analysis of the dynamics of trajectories near heteroclinic cycles and networks is often done via construction of Poincar\'e maps to approximate the flow (see, e.g.~\cite{KS94}). The return maps are an ensemble of maps where each is a composition of two types of map: local maps past neighbourhoods of the equilibria and global map that connect the neighbourhoods. The local maps are constructed assuming no resonances and then linearising the flow near the equilibria while the global maps are truncated Taylor expansions of the maps near the connecting orbits. The lowest order truncations of these maps then approximate the flow close to the heteroclinic network.

The detailed construction of such maps for the networks (even in the noise-free case) is beyond the scope of this paper. However, in later sections we apply previous results regarding the dynamics of trajectories near heteroclinic networks in the presence of noise, so we find it useful to define the appropriate Poincar\'e sections here.
Poincar\'e sections are often defined to be surfaces a distance $h$ from an equilibrium, where $h$ is some small constant, meaning they are spheres. However, in order to apply results of Stone, Armbruster and others~\cite{StoHol90,StoArm99,ArmStoKir03} we define  Poincar\'e sections as unions of codimension one surfaces as follows. We give explicit definitions of Poincar\'e sections for the simplex realisation; for the cylinder realisation they can be defined similarly.

Consider an equilibrium  $\xi_k$ in the simplex realisation, with an incoming heteroclinic connection from equilibria $\xi_{j}$, and an outgoing heteroclinic connection towards equilibria $\xi_{l}$. We define (for a fixed small $h$):
\begin{equation} \label{eq:Hin}
\Hin[,j]{k}=\{x\in\R^{n_v} : x_{j}=h, |1-x_k|<h,  |x_i|<h, i\neq j,k \}
\end{equation}
\begin{equation} \label{eq:Hout}
\Hout[,l]{k}=\{x\in\R^{n_v} :  x_{l}=h, |1-x_k|<h, |x_i|<h, i\neq l,k \}.
\end{equation}

\begin{figure}
\psfrag{xi1}{$\xi_1$}
\psfrag{xi2}{$\xi_2$}
\psfrag{xi3}{$\xi_4$}
\psfrag{xi4}{$\xi_3$}
\psfrag{x1}{}
\psfrag{x3}{}
\psfrag{x4}{}
\psfrag{H2in}{$\Hin[,1]{2}$}
\psfrag{H2out}{$\Hout[,4]{2}$}
\psfrag{H2out4}{$\Hout[,3]{2}$}
\begin{center}
\epsfig{file=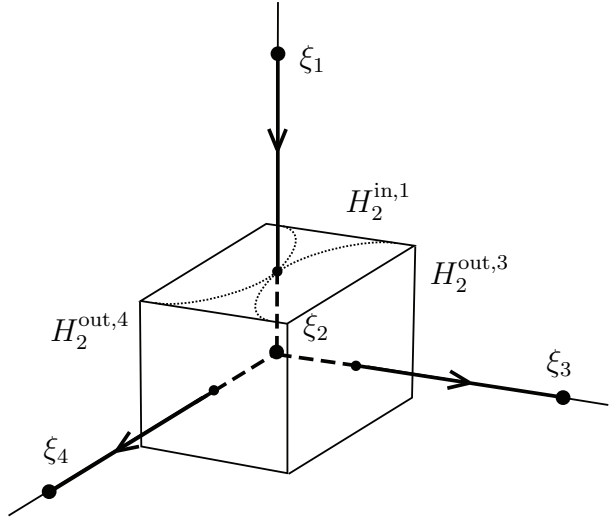,width=8cm}
\caption{\label{fig:Psects} Schematic diagram showing a Poincar\'e sections $\Hin[,1]{2}$, $\Hout[,3]{2}$ and $\Hout[,4]{2}$  near the equilibrium $\xi_2$. Heteroclinic orbits are shown with bold lines. Large dots indicate equilibria, small dots indicate the intersection of the heteroclinic orbits with the Poincar\'e sections. The dotted lines on $\Hin{2}$ indicate the dividing line between regions of trajectories which will next travel to $\xi_3$ or $\xi_4$, in the case that $e_{23}>e_{24}$. 
}
\end{center}
\end{figure}

If a vertex in the graph has an incoming or outgoing degree greater than one, then there will be multiple incoming or outgoing Poincar\'e sections, and so we also define
\[
\Hin{k}=\bigcup_j \Hin[,j]{k},\quad \Hout{k}=\bigcup_l \Hout[,l]{k}
\]
where the unions are taken over all incoming and outgoing directions respectively. Note that if the trajectory spends time near $\xi_k$ it must have passed through $\Hin{k}$ and will pass through $\Hout{k}$ at some time thereafter. Figure~\ref{fig:Psects} shows a schematic diagram of the Poincar\'e sections near an equilibrium $\xi_2$, with an incoming connection from $\xi_1$ and outgoing connections to $\xi_3$ and $\xi_4$. 

In the following sections, we use the following notation for eigenvalues of equilibria. If $\xi_k$ has a contracting direction in the $x_j$ direction, we label the corresponding eigenvalue $-c_{kj}$. Similarly, if  If $\xi_k$ has an expanding direction in the $x_l$ direction, we label the corresponding eigenvalue $e_{kl}$.

\subsection{Transition probabilities between equilibria}
\label{sec:trans_prob}

If $\xi_j$ has a single expanding direction towards $\xi_k$, then clearly $\pi_{j,k}=1$ and $\pi_{j,\ell}=0$ for $k\neq \ell$ in the limit of low noise. The more interesting case is when the equilibrium has two expanding directions, as in the example shown schematically in Figure~\ref{fig:Psects}. Here, at the equilibrium $\xi_2$, the trajectory makes a `choice' as to whether to next visit $\xi_3$ or $\xi_4$. We give here an outline of how to compute the probabilities of the trajectory making each choice; that is, the transition probabilities.
 The extension of these calculations to equilibria with three or more expanding directions is straightforward.

Armbruster et.\ al~\cite{ArmStoKir03} compute the transition probabilities for an equilibrium with two expanding directions, under the assumption that the incoming distribution of coordinates of a trajectory is approximately Gaussian and centered at zero. For the schematic example in Figure~\ref{fig:Psects}, we show this distribution of incoming trajectories, in terms of the $x_3$ and $x_4$ coordinates on $\Hin{2}$, on the left-hand side of Figure~\ref{fig:H2in}. The proportion of trajectories which we expect to visit $\xi_4$ is given by the proportion of the measure in the `noise ellipse' which intersects the cusp.

The computation of this area was given in~\cite{ArmStoKir03}. The results depend on constants which come from the global part of the flow and are in general unknown. However, a scaling can be found in the limit of low noise. It was shown that if an equilibrium $\xi_j$ has expanding directions $x_k$ and $x_m$, with expanding eigenvalues $e_{jk}>e_{jm}$, then in the limit of low noise, (i.e.\ as $\zeta\rightarrow 0$) the transition probability from $\xi_j$ to $\xi_m$ scales as
\[
\pi_{j,m}=O\left(\zeta^{\frac{e_{jk}}{e_{jm}}-1}\right).
\]

\subsection{Lift-off and memory effects for transitions between equilibria}
\label{sec:liftoff}

For some applications, it may be desirable for a network to have memory effects, i.e. where the transition probabilities depend on the recent history of the trajectory taken through the network. It is possible to create memory effects in a noisy heteroclinic network that has ``lift-off'', an effect noted by Stone, Armbruster and Kirk~\cite{StoArm99,ArmStoKir03} that appears for attracting heteroclinic (but not homoclinic) networks with low amplitude additive noise.

\begin{figure}
\begin{center}
\epsfig{file=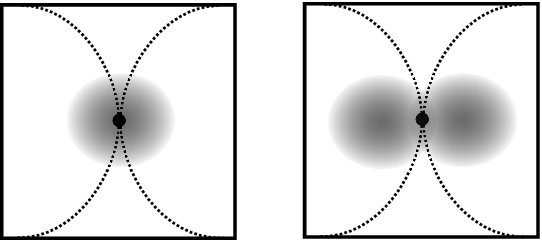,width=12cm}
\caption{\label{fig:H2in} Schematic diagram showing the Poincar\'e section $\Hin[,1]{2}$ for the connection from $\xi_1$ to $\xi_2$ shown in Figure~\ref{fig:Psects}, projected onto the $x_3,x_4$ plane. The dotted lines indicate the dividing line between regions of trajectories which travel next to $\xi_3$ or $\xi_4$. The shaded ellipse represents the distribution of coordinates of trajectories as they pass through the section where the darker regions are visited more often, while the black dot shows the location of the noise-free connection. The left panel shows a case in which the incoming distribution of the $x_3$ and $x_4$ coordinates is approximately Gaussian and centered at zero. The right panel shows a case with lift-off in the $x_3$ direction and consequently a multi-modal distribution.  }
\end{center}
\end{figure}

Lift-off is a property of the distribution of the coordinates of a trajectory as it enters a neighborhood of an equilibrium such that the probability distribution of coordinates of trajectories in a Poincar\'{e} section becomes multi-modal and is a mechanism by which the transition probabilities can gain memory. We demonstrate the idea of lift-off in Figure~\ref{fig:H2in}, which shows the Poincar\'e section $\Hin[,1]{2}$ for the schematic in Figure~\ref{fig:Psects}, and a representation of the distribution of the coordinates of the incoming trajectories.  If no lift-off occurs, then the distribution of coordinates is approximately Gaussian and centered at zero, as shown on the left. If lift-off occurs, then the distribution of a particular coordinate is no longer Gaussian and may include several peaks, as shown on the right. 

The transition probabilities can be thought of as intersections of the distribution of incoming trajectories with the set of points that progress to a given equilibrium at the next epoch. As lift-off changes this area of intersection and depends on the previously epochs, lift-off can affect transition probabilities and make them depend on memory. The occurrence of lift-off depends on the eigenvalues of the equilibria, and hence can be controlled. Note that a sufficient condition for lift-off \emph{not} to occur is that all contracting eigenvalues are greater than all expanding eigenvalues. 

It was shown in~\cite{StoArm99} that if the network contains a heteroclinic connection $\xi_j\rightarrow\xi_k$, then there is the possibility of creating lift-off in the $x_j$ direction as the trajectory exits a neighbourhood of the equilibrium $\xi_k$. If this lift-off can be maintained until the trajectory next approaches an equilibrium with an unstable manifold which connects to $\xi_j$, the probability of visiting $\xi_j$ will be higher than if the lift-off had not occurred. 

Stone and Armburster~\cite{StoArm99} compute conditions on the eigenvalues for lift-off to occur. They show that, if there exists a  connection $\xi_j\rightarrow\xi_k$, lift-off will occur at $\xi_k$ (that is, in the distribution of coordinates of trajectories leaving a neighbourhood of $\xi_k$) in the $x_j$ direction if 
$$
\frac{c_{kj}}{e_{kl}}<1,
$$
where $x_l$ is the expanding direction at $\xi_k$. That is, if this condition is satisfied, then the distribution of the $x_j$ coordinate, as the trajectory exits a neighbourhood of $\xi_k$, is not centered about zero but instead is centered about a point whose distance from the connection scales as
\[
\zeta^{\frac{c_{kj}}{e_{kl}}}.
\]
in the low noise limit $\zeta\rightarrow 0$. Since the distribution of the noise is symmetric about zero, the lift-off has equal probability of occurring in either the positive or negative direction. This means that the distribution becomes multimodal.
Since the global flow only scales the distribution of coordinates by an amount that is order one in $\zeta$, the incoming distribution to the next equilibrium will scale similarly.

It is straightforward to extend the calculations of~\cite{StoArm99} to a sequence of heteroclinic connections between equilibria $\xi_j\rightarrow\xi_k\rightarrow\xi_l\rightarrow\xi_m$. That is, we already have conditions that the distribution of the $x_j$ coordinate as it exits a neighbourhood of $\xi_k$ is not centered about zero, and we can extend the computations to give conditions that the distribution of the $x_j$ coordinate as the trajectory exits a neighbourhood of $\xi_l$ is similarly not centered about zero. 

These calculations are a simple extension of those given in~\cite{StoArm99}, and it can be shown that 
\begin{equation}\label{eq:lo2}
\frac{c_{kj}}{e_{kl}}+\frac{c_{lj}}{e_{lm}}<1,
\end{equation}
implies that the distribution of the $x_j$ coordinate as the trajectory exits a neighbourhood of $\xi_l$ is centered about a point whose distance from the connection scales as
\[
\zeta^{\frac{c_{kj}}{e_{kl}}+\frac{c_{lj}}{e_{lm}}}.
\]
in low noise limit $\zeta\rightarrow 0$. Again, since the global flow only scales coordinates by an order one amount, the incoming distribution of the $x_j$ coordinate at equilibrium $\xi_m$ will then also not be centered at zero. Thus, if condition~\eqref{eq:lo2} on the eigenvalues is satisfied, and $\xi_m$ has an unstable manifold (of dimension two or more) which includes the direction towards $\xi_j$, then the proportion of trajectories which go towards $\xi_j$ will be higher than if lift-off had not occurred.

Note that lift-off occurs only if the contracting eigenvalues are small enough, that is, if there is not enough contraction in the $x_j$ direction at $\xi_l$ to `squash' the lift-off. Further extension of these calculations will give conditions on lift-off being maintained over longer sequences of equilibria, and hence the possibility of longer term memory. However, as lift-off requires sufficiently small contracting eigenvalues ``on average'', and stability of the network requires sufficiently large contracting eigenvalues, we conjecture that any memory effects present must appear of a sequence of epochs that is smaller than the longest cycle within the network.

\section{Example: simplex realisation of the decision graph}
\label{sec:statsnumerics}

In this section, we numerically investigate the properties described in Section~\ref{sec:stats} for the decision graph realised using the simplex realisation of Section~\ref{sec:decision}. Note that whilst it is comparatively easy to check for the presence of memory in solution trajectories, checking for the \emph{absence} of memory is difficult. That is, although we can perform statistical tests to show that a trajectory may be zeroth order rather than first order Markov, checking for the absence of all long-term memory effects is hard. Whilst we expect that such long term memory (i.e. over sequences of more than two equilibria) is possible, it seems unlikely to occur in the absence of short term memory except in special cases. A detailed study of long-term memory is beyond the scope of this paper and so in the following sections we discuss only the presence or absence of short-term memory.

\subsection{Transition probabilities in the decision graph}

First, we consider a particular case where we predict that the network is memoryless, that is, there is no lift-off, and so the incoming distributions of all coordinates at each equilibria are Gaussians centered at zero. 

We confirm the scaling given in Section~\ref{sec:trans_prob} for the transition probabilities in the
decision graph by performing numerical integrations at various noise levels. We use parameters
$$
\begin{array}{c}
c_{jk}=t_{jk}=2, e_{12}=2, e_{24}=e_{36}=e_{48}=1.85,\\
e_{23}=e_{35}=e_{47}=1.99, e_{51}=e_{61}=e_{71}=e_{81}=1.98
\end{array}
$$
and integrate for noise levels $\zeta=10^{-5}, 10^{-6}, 10^{-7}, 10^{-8}, 10^{-9}, 10^{-10}, 10^{-11}$. For each noise level we integrate until the number of passes the trajectory makes through $\xi_1$ is $5000$. 

\begin{figure}
\psfrag{logeps}{\raisebox{-0.2cm}{$\log\zeta$}}
\psfrag{logprop}{\hspace{-0.7cm}$\log$ (proportion)}
\begin{center}
\epsfig{file=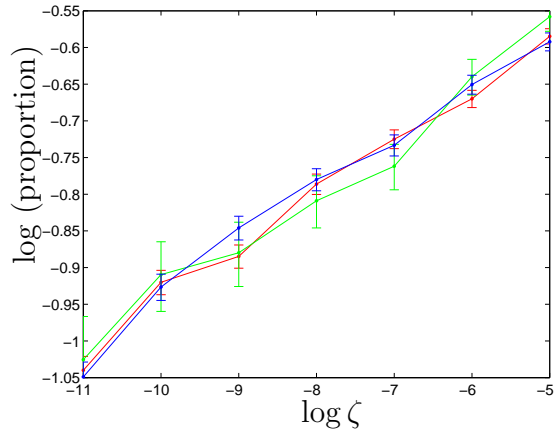,width=8cm}
\end{center} 
\caption{\label{fig:props_noise} 
For the decision graph and simplex realisation, the figure shows number of visits the trajectory makes to the equilibria $\xi_4$ (red), $\xi_6$ (blue) and $\xi_8$ (green), as a proportion of the number of visits to $\xi_2$, $\xi_3$ and $\xi_4$ respectively, for various noise levels $\zeta$; the bars show standard errors. For details and parameter values, see text.}
\end{figure}

At each noise level, we measure the number of visits the trajectory makes to the equilibria $\xi_4$, $\xi_6$ and $\xi_8$, as a proportion of the number of visits to $\xi_2$, $\xi_3$ and $\xi_4$ respectively.  In terms of the definitions given in Section~\ref{sec:stats}, these are approximations of
\[
\frac{\pi_4}{\pi_2},\quad \frac{\pi_6}{\pi_3},\quad \mathrm{and}\quad \frac{\pi_8}{\pi_4}
\]
respectively, using approximations (\ref{eq:approxprob}) and $k=5000\times 4=20,000$ with standard errors. These proportions are plotted against the noise level in Figure~\ref{fig:props_noise} on log-log axes.  Best fit linear regressions for each of these lines have slopes 0.072, 0.074 and 0.073 respectively; note that the expected slope for all three is $\frac{1.99}{1.85}-1=0.076$. This means that in the low noise limit $\zeta\rightarrow 0$ we predict that almost all itineraries follow the cycle 
$$
\cdots\rightarrow \xi_1\rightarrow\xi_2\rightarrow\xi_3\rightarrow \xi_5\rightarrow \xi_1\rightarrow \cdots
$$
corresponding to simply selecting the most unstable expanding direction at each equilibrium.

Note that without knowing details about the global part of the flow, it is not possible to predict the precise transition probabilities for particular noise levels and eigenvalues, the best we can do is get the scaling for low noise. However, if a particular transition probability is desired, a little experimentation and alteration of noise levels can achieve this.

\subsection{Numerical examples of memory effects for the decision graph}

We now demonstrate an example of how lift-off as described in Section~\ref{sec:liftoff} can create memory in a network. We use the decision graph with the simplex realisation as an example, and induce lift-off in the $x_3$ direction at $\xi_5$. We give two examples: in the first the lift-off is maintained until the trajectory reaches $\xi_2$. In this example, we expect that trajectories which visit $\xi_5$ will then have a larger probability of visiting $\xi_3$ on the next circuit of the network than they would have done had they previously visited $\xi_6$, $\xi_7$ or $\xi_8$. In the second example, we exhibit parameters such that the lift-off is not maintained at $\xi_2$, and consequently there are no memory effects.

The particular structure of the decision graph can be used to simplify our analysis. Note that the sequence of equilibria visited (that is, the sequence $i_k, k=1,\ldots$) can be deduced by only recording which of the equilibria  $\xi_5$, $\xi_6$, $\xi_7$ and $\xi_8$ was visited by the trajectory on each `loop' around the network, and we therefore focus just on transitions between this subset of equilibria. 

Specifically, we consider the sub-itinerary $\tilde{\iota}_k$ of (\ref{eq:itinerary}) that consist of visits to 
$\{5,6,7,8\}$ by choosing the minimal strictly increasing set of indices $0<\ell_k$ such that
\[
\tilde{\iota}_k=i_{\ell_k}\in \{5,6,7,8\}.
\]
Observe that from the structure of the decision graph in Figure~\ref{fig:decisiongraph} we expect that $\ell_{k+1}-\ell_k=4$ on most occasions, and note also that we can have one-cycles (non-trivial transitions from $j$ straight back to $j$) in this induced graph. Using the corresponding definitions for proportions of visits and transition probabilities between this subset of equilibria we define for any sequence of $j_k\in\{5,6,7,8\}$:
\begin{equation}\label{eq:tildefiprob}
\tcP(j_1,\ldots,j_m)=\mP(~(\tilde{x},\tilde{\vartheta})~:~\tilde{\iota}_{\ell}=j_\ell~\mbox{ for }~\ell=1,\ldots,m \mbox{, for the trajectory starting at $(\tilde{x},\tilde{\vartheta})$}).
\end{equation}
Analogously to (\ref{eq:approxprob}) we can compute this from a typical trajectory as
\begin{equation}\label{eq:tildeapproxprob}
\tcP(j_1,\ldots,j_m)=\lim_{k\rightarrow \infty} \frac{1}{k} \#\{ 0 \leq \ell < k ~:~ \tilde{\iota}_{\ell+n}=j_n \mbox{ for }n=1,\ldots,m\}
\end{equation}
and so define the probability $\tilde{\pi}_j$ of the first epoch being close to $j\in\{5,6,7,8\}$:
$$
\tilde{\pi}_j= \tcP(j).
$$
Assuming $\tilde{\pi}_j>0$, we define the transition probability between vertices $\xi_{j_1}$ and $\xi_{j_2}$ by:
\begin{equation}\label{eq:tildetprob}
\tilde{\pi}_{j_1,j_2}=\frac{\tcP(j_1,j_2)}{\tilde{\pi}_{j_1}}.
\end{equation}

We perform numerical integrations using the Heun method (with timestep $0.01$) for the cases with and without memory, and observe the effects in two ways. First, we compute the observed number of transitions between this subset of equilibria to estimate the matrix of probabilities $\tilde{\pi}_{j_1,j_2}$ for $j_1,j_2=5,\dots,8$.  A $\chi$-squared test is used to determine whether or not the probabilities of visiting each of the equilibria are independent of the previous state of the trajectory. We also directly observe the distribution of coordinates as trajectories enter and leave neighbourhoods of equilibria. From the results in Section~\ref{sec:liftoff}, we know that lift-off at $\xi_5$ in the $x_3$ direction occurs if $\frac{c_{53}}{e_{51}}<1$. This lift-off will still exist after the trajectory when the trajectory enters a neighbourhood of $\xi_2$ if $\frac{c_{53}}{e_{51}}+\frac{c_{13}}{e_{12}}<1$. We first perform an experiment where both conditions are satisfied, and then an experiment where only the first condition is satisfied.

\subsection{An example with memory}
\label{sec:ex_mem}

We choose parameters so that
\[
\frac{c_{53}}{e_{51}}<1 \quad \mathrm{and}\quad \frac{c_{53}}{e_{51}}+\frac{c_{13}}{e_{12}}<1,
\]
and integrate the equations for the decision graph simplex realisation, with noise $\zeta=10^{-5}$, and for total time $200,000$, giving a sequence of $31,568$ epochs and so $k_c=31,568/4$ cycles. The parameters used are
$c_{jk}=2$ for all $j$, $k$ except for $c_{13}=0.8$, $c_{53}=0.8$, $t_{jk}=2$ for all $j$, $k$, $e_{12}=2$, $e_{24}=e_{36}=e_{48}=1.9$, $e_{23}=e_{35}=e_{47}=1.99$, and $e_{51}=e_{61}=e_{71}=e_{81}=1.98$.

For a randomly chosen initial condition and noise path, the following matrix is the observed number of transitions between equilibria $\xi_5$, $\xi_6$, $\xi_7$ and $\xi_8$ in the reduced system:
\[
\begin{pmatrix}
2829 & 1478 &143 & 76 \\
1166 & 643 & 379 & 181   \\
 355 & 158 & 88 & 57\\
 176& 90 &48 & 25  \\
\end{pmatrix}
\]
which gives an approximation to the matrix $\tilde{\pi}_{j_1,j_2}$:
\[
\begin{pmatrix}
\tpi_{5,5} & \tpi_{5,6} & \tpi_{5,7} & \tpi_{5,8} \\
\tpi_{6,5} & \tpi_{6,6} & \tpi_{6,7} & \tpi_{6,8} \\
\tpi_{7,5} & \tpi_{7,6} & \tpi_{7,7} & \tpi_{7,8} \\
\tpi_{8,5} & \tpi_{8,6} & \tpi_{8,7} & \tpi_{8,8}
\end{pmatrix}
\approx
\begin{pmatrix}
0.6251 & 0.3266 & 0.0316  & 0.0168 \\
    0.4922  &  0.2714  &  0.1600 &   0.0764 \\ 
       0.5395 &   0.2401 &   0.1337  &  0.0866 \\
    0.5192  & 0.2655 &   0.1416  &  0.0737
\end{pmatrix}
\]
where standard errors are of order $1/\sqrt{k_c}$, i.e. $\pm 0.01$. Hence, if a trajectory visits $\xi_5$, it is more likely to go via $\xi_3$ (and hence to $\xi_5$ or $\xi_6$ rather than $\xi_7$ or $\xi_8$) on its next circuit around the network. A $\chi$-squared test confirms this by rejection the null hypothesis that the transition probabilities are independent of the previous equilibrium visited.

\begin{figure}
\psfrag{x1}{\raisebox{-0.1cm}{$x_1$}}
\psfrag{x3}{\raisebox{-0.1cm}{$x_3$}}
\begin{center}
\subfigure[Distribution of $x_1$ on $\Hin{5}$]{\epsfig{file=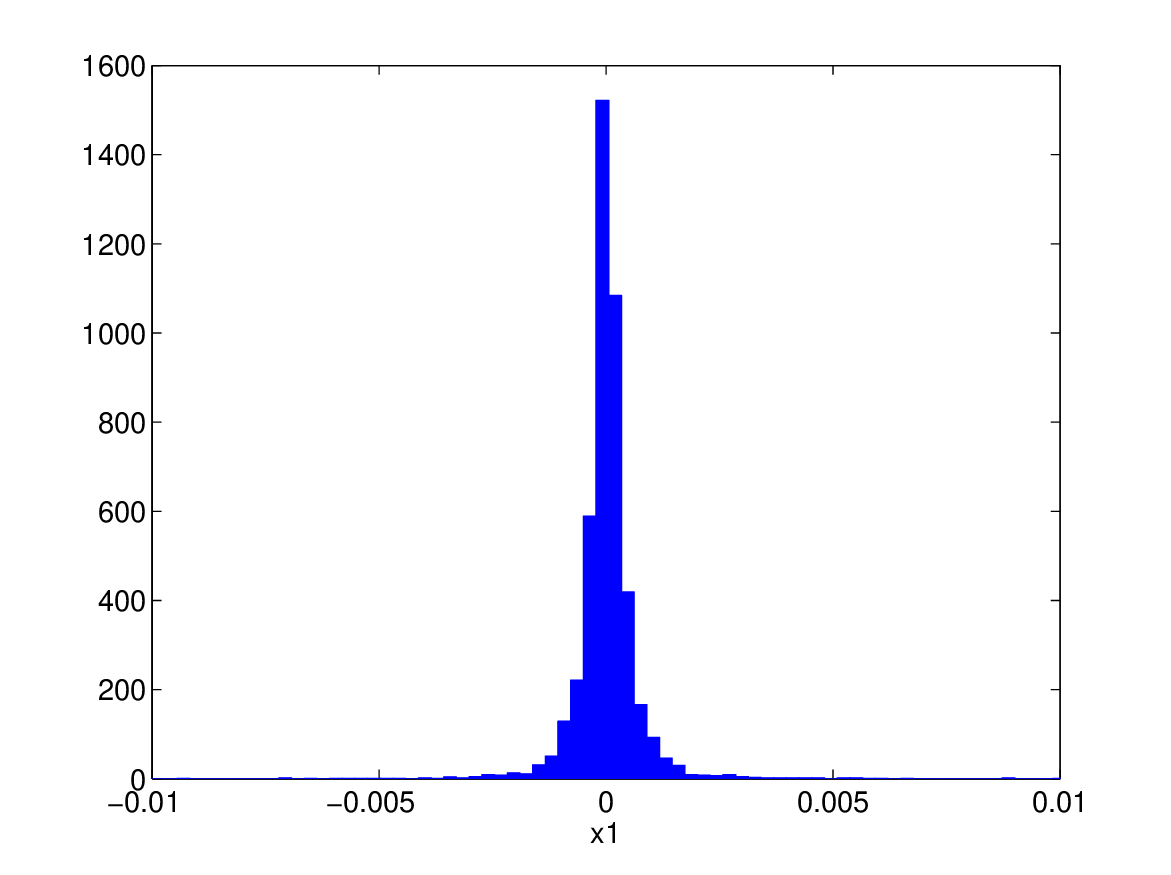,width=7cm}} 
\subfigure[Distribution of $x_3$ on $\Hout{5}$]{\epsfig{file=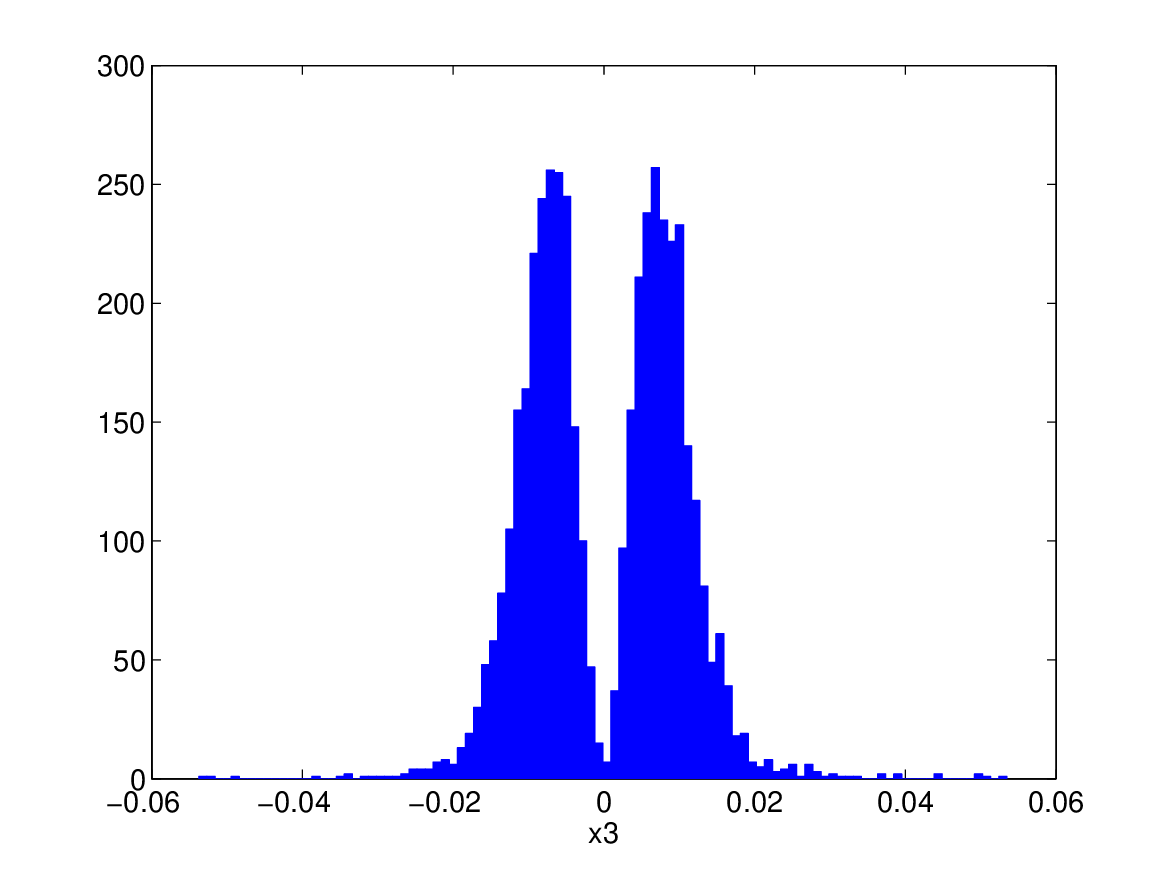,width=7cm}} \\
\subfigure[Distribution of $x_3$ on $\Hout{1}$]{\epsfig{file=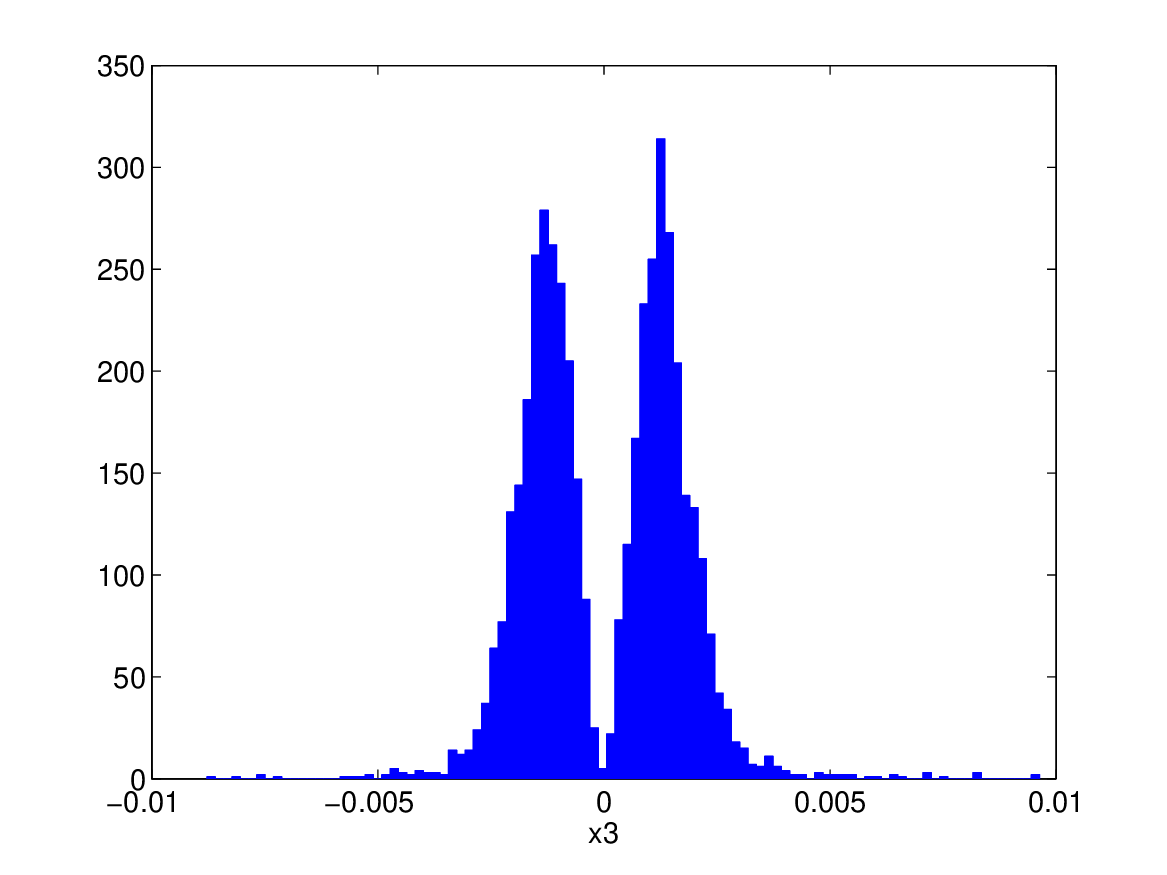,width=7cm}} 
\subfigure[Distribution of $x_3$ on $\Hout{1}$]{\epsfig{file=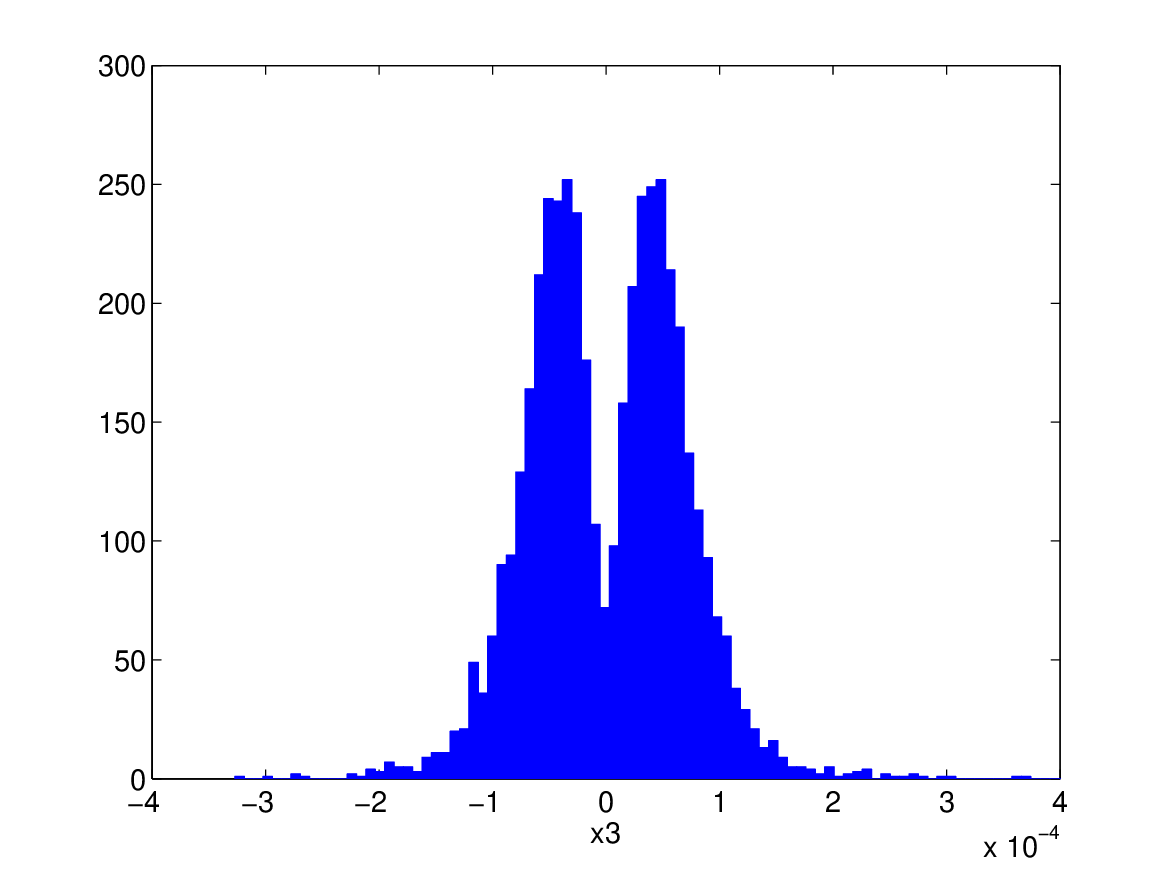,width=7cm}}
\end{center} 
\caption{\label{fig:x4_liftoff} Distributions of coordinates for trajectories which pass from equilibrium $\xi_3$ to $\xi_5$ to $\xi_1$, in the decision graph. Parameters satisfy $\frac{c_{53}}{e_{51}}+\frac{c_{13}}{e_{12}}<1$, and so lift-off in the $x_3$ coordinate is expected and can be seen in the non-Gaussian distributions.}
\end{figure}

Figure~\ref{fig:x4_liftoff} demonstrates the lift-off by showing distributions of the $x_1$ and $x_3$ coordinates as the trajectory passes from equilibria $\xi_5$ to $\xi_1$. We plot the coordinates as the trajectory crosses appropriate Poincar\'e sections: $\Hin{5}$, $\Hout{5}$, $\Hin{1}$ and $\Hout{1}$, using $h=0.1$. On $\Hin{5}$, $x_3$ is of order $h$, and the coordinate of interest is $x_1$. Figure~\ref{fig:x4_liftoff}(a) shows this to clearly be approximately Gaussian centered at zero. Figures~\ref{fig:x4_liftoff}(b) to (d) show distributions of the $x_3$ coordinate on $\Hout{5}$, $\Hin{1}$ and $\Hout{1}$.  In each of these three cases, the distribution is clearly not Gaussian, but instead is approximately the sum of two Gaussians shifted to the right and left of zero. This demonstrates the lift-off of the trajectory both in the positive and negative directions, on different loops around the network.
 
\begin{figure}
\psfrag{x4}{$x_4$}
\psfrag{x3}{\raisebox{-0.2cm}{$x_3$}}
\begin{center}
\epsfig{file=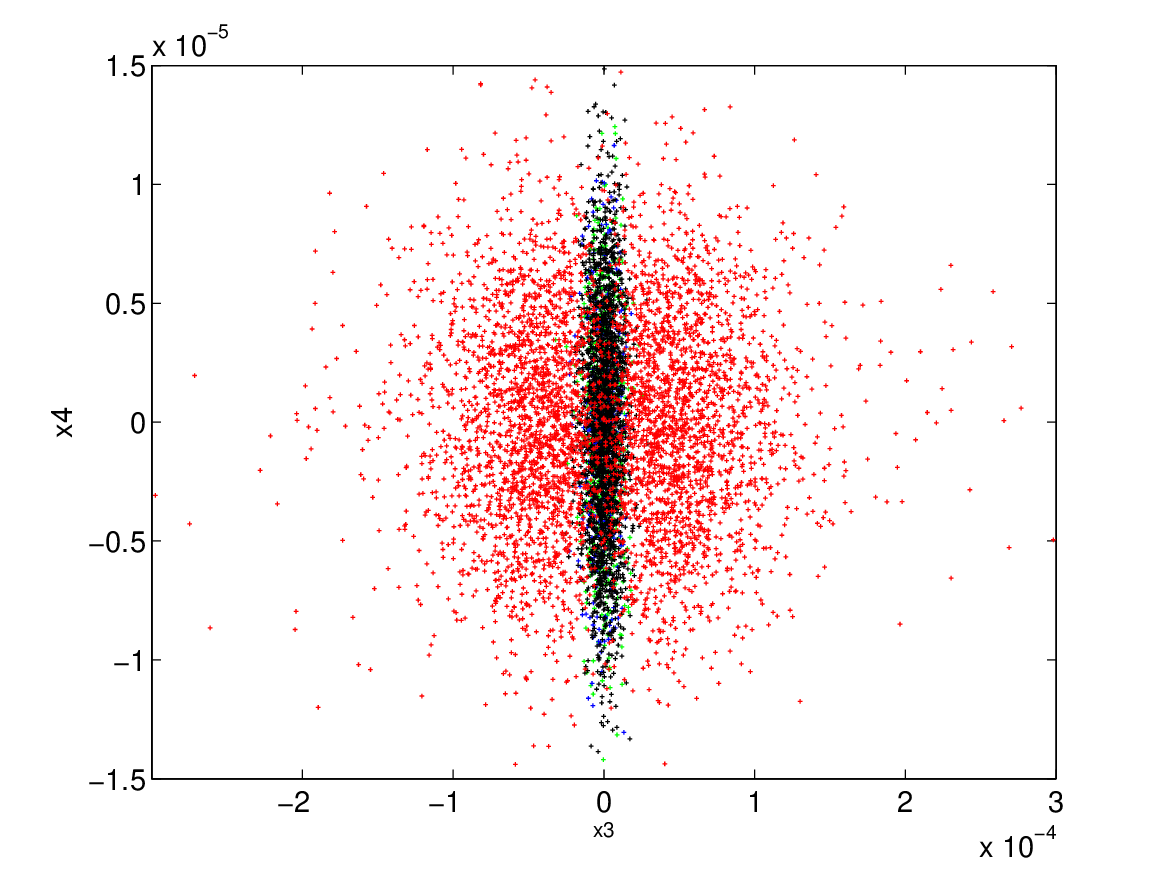,width=9cm}
\end{center} 
\caption{\label{fig:all_scatter}  Scatter plot of $x_3$ and $x_4$ coordinates showing the condition distribution of trajectories as they exit from $\xi_1$ on $\Hout{1}$, for the same trajectory plotted in Figure~\ref{fig:x4_liftoff}. Trajectories which visited $\xi_5$ on the previous loop around the network are coloured red, while trajectories which previously visited $\xi_6$, $\xi_7$ and $\xi_8$ are coloured black, blue and green (note that the blue and green points are plotted under the black points, making them difficult to see). Lift-off in the $x_3$ direction for those points that are coloured red can clearly be seen to cause memory.}
\end{figure}

The conditional distributions in Figure~\ref{fig:all_scatter} clearly show the memory effect via a scatter plot of the $x_3$ and $x_4$ coordinates of the trajectory on $\Hout{1}$, conditional on the previously visited cycle. The conditional distribution peaks are clearly different as only those that went past $\xi_6$ at the last cycle display lift-off.

\subsection{An example with no memory}

We now explore a second example with same parameters as in Section~\ref{sec:ex_mem}, except that $c_{13}=1.3$, so 
\[
\frac{c_{53}}{e_{51}}<1 \quad \mathrm{but}\quad \frac{c_{53}}{e_{51}}+\frac{c_{13}}{e_{12}}>1.
\]
This means we expect to see lift-off in the $x_3$ direction at $\xi_5$, but after the trajectory has passed $\xi_1$ this lift-off will be `squashed' and hence the distribution of the $x_3$ coordinate as the trajectory enters a neighbourhood of $\xi_2$ will be Gaussian again. Thus the network will have no memory.
Again, we integrate for time $200,000$, which gives a sequence of $k_c=30,892/4$ cycles. 

For an example run, the observed number of transitions between equilibria $\xi_5$, $\xi_6$, $\xi_7$, and $\xi_8$ is:
\[
\begin{pmatrix}
1686     &    957  &       660  &       348 \\
         896 &        446 &        371 &        186 \\
          697    &     328    &     254    &     142 \\
         373       &  167       &  136      &    76 \\
\end{pmatrix}
\]
which gives the following approximation to $\tpi_{j_1,j_2}$:
\[
\begin{pmatrix}
\tpi_{5,5} & \tpi_{5,6} & \tpi_{5,7} & \tpi_{5,8} \\
\tpi_{6,5} & \tpi_{6,6} & \tpi_{6,7} & \tpi_{6,8} \\
\tpi_{7,5} & \tpi_{7,6} & \tpi_{7,7} & \tpi_{7,8} \\
\tpi_{8,5} & \tpi_{8,6} & \tpi_{8,7} & \tpi_{8,8}
\end{pmatrix}
\approx
\begin{pmatrix}
0.4618 &   0.2621 &   0.1808 &   0.0953 \\
    0.4718 &   0.2349 &   0.1954 &   0.0979 \\
    0.4905 &   0.2308  &  0.1787 &   0.0999 \\
    0.4960 &   0.2221  &  0.1809 &   0.1011 \\
\end{pmatrix}
\]
with standard errors of order $\pm 0.01$. In this case the $\chi$-squared test does not reject the null hypothesis of independence for this transition matrix.

\begin{figure}
\psfrag{x3}{\raisebox{-0.1cm}{$x_3$}}
\begin{center}
\subfigure[Distribution of $x_3$ on $\Hin{1}$]{\epsfig{file=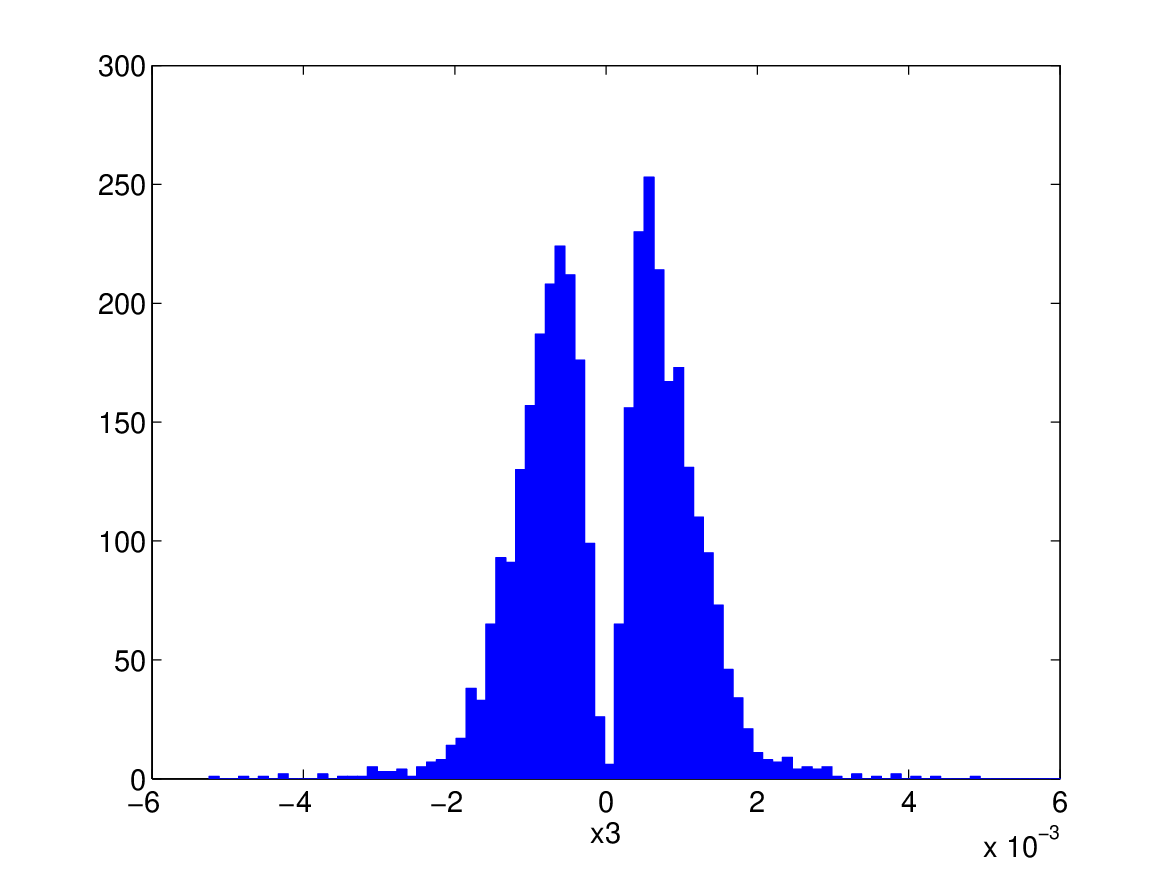,width=7cm}} 
\subfigure[Distribution of $x_3$ on $\Hout{1}$]{\epsfig{file=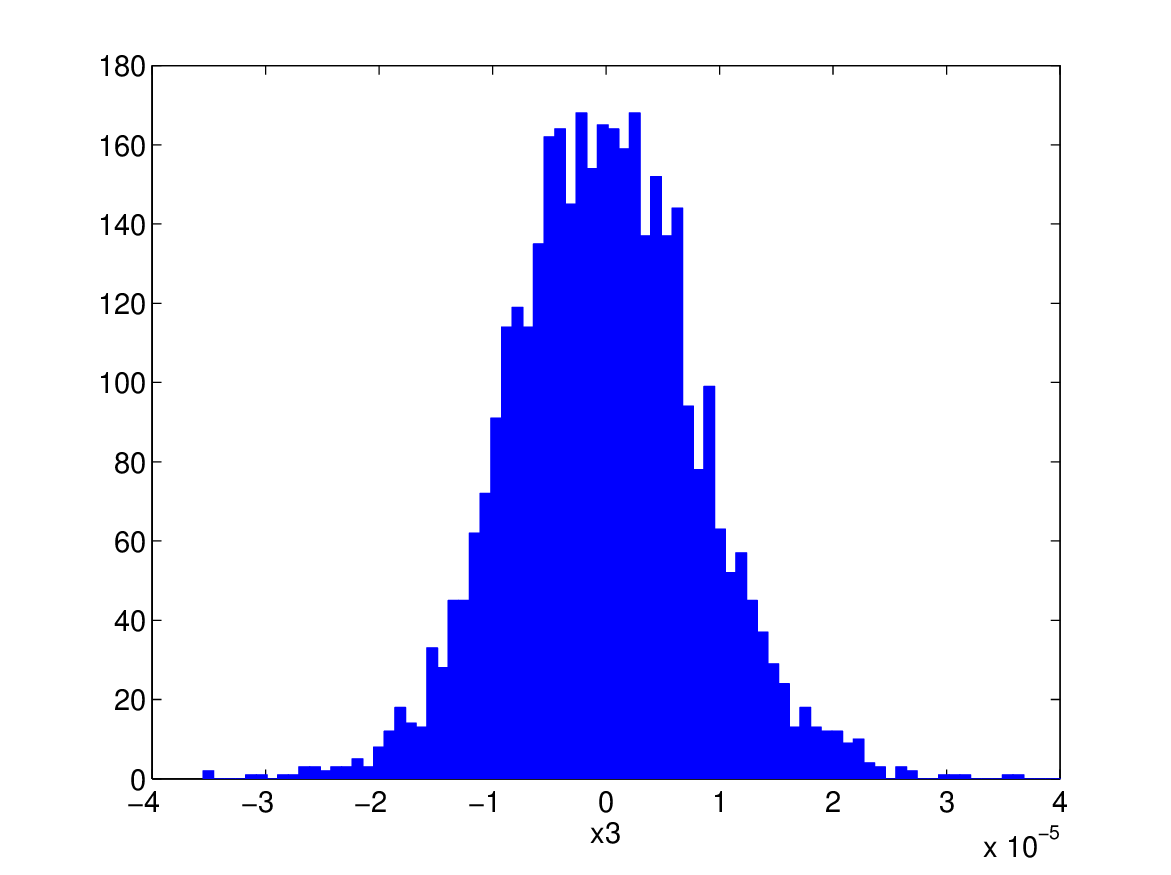,width=7cm}}
\end{center} 
\caption{\label{fig:x4_noliftoff} Distributions of coordinates for trajectories crossing $\Hin{1}$ and $\Hout{1}$, having previously visited $\xi_5$. Parameters in this case satisfy $\frac{c_{53}}{e_{51}}<1$ but  $\frac{c_{53}}{e_{51}}+\frac{c_{13}}{e_{12}}>1$. Lift-off in the $x_3$ direction can be seen as the trajectory enters a neighbourhood of $\xi_1$ but it has been compressed by the time the trajectory leaves a neighbourhood of $\xi_1$.}
\end{figure}

\begin{figure}
\psfrag{x4}{$x_4$}
\psfrag{x3}{\raisebox{-0.2cm}{$x_3$}}
\begin{center}
\epsfig{file=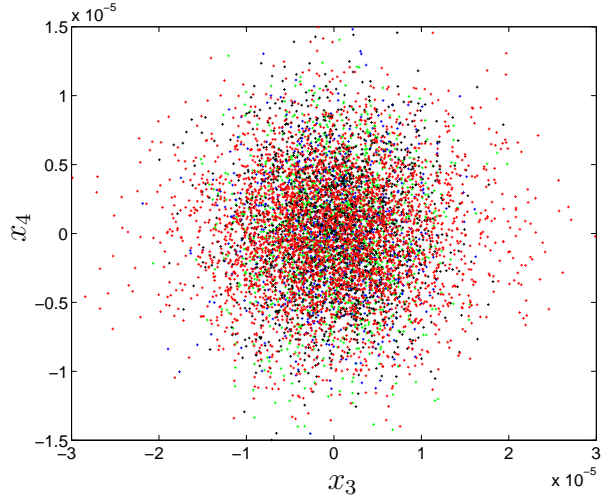,width=9cm}
\end{center} 
\caption{\label{fig:all_scatter_no_lo} Scatter plot of $x_3$ and $x_4$ coordinates showing the condition distribution of trajectories as they exit from $\xi_1$ on $\Hout{1}$, for the same trajectory plotted in Figure~\ref{fig:x4_noliftoff}. Trajectories which visited $\xi_5$ on the previous loop around the network are coloured red, while trajectories which previously visited $\xi_6$, $\xi_7$ and $\xi_8$ are coloured black, blue and green. The distributions of all four sets of trajectories appears to be identical, indicating that there is no memory. 
}
\end{figure}

Figure~\ref{fig:x4_noliftoff} shows the distribution of the $x_3$ coordinate (for those trajectories which pass through $\xi_5$) on $\Hin{1}$ and $\Hout{1}$. It can clearly be seen that there is lift-off in the $x_3$ direction before $\xi_1$, but on
exiting, the distribution has returned to being approximately Gaussian with zero mean; this is shown as a scatter plot in Figure~\ref{fig:all_scatter_no_lo} where, in contrast to the case in Figure~\ref{fig:all_scatter}, the distributions do not appear to depend on previous epochs and so the system is memoryless.

\section{Discussion}
\label{sec:discuss}

We have shown that, in principle, any finite directed graph can be realised as an embedded attracting robust heteroclinic network for a coupled cell system. In fact the method should work even if the network is not connected (respectively, if the components are not strongly connected) - in these cases the resulting network will be disconnected (respectively, the asymptotic dynamics will be contained within a subnetwork). There are clearly issues at vertices of high degree for outgoing edges - there must be nontrivial dynamics such as additional saddle points involved in separating trajectories that go to different edges. We have not been able to prove all we would like to concerning the dynamics of the constructed networks; in particular we have not been able to characterize the larger embedding network $\tilde{X}$ for either simplex or cylinder realisations. Nonetheless, our numerical examples suggest that the dynamics is robust enough to be useful as a method for designing dynamical systems with specific graph attractors in phase space.

The noisy networks will be qualitatively robust to perturbations of parameters up to the point where the network undergoes one of a number of possible bifurcations. As suggested by \cite[Figures 14 and 15]{AshOroWorTow07}, one bifurcation that will destroy the heteroclinic network and replace it with an excitable network is when one of the vertices becomes locally stable via a pitchfork bifurcation. At such a pitchfork bifurcation, new saddles will be created whose stable manifolds give excitability thresholds - the connections no longer form a chain recurrent set. Another bifurcation will be a resonance that causes the network to lose stability and results in bifurcation of periodic or other attractors from the network. Unlike the heteroclinic networks considered in \cite{dias_etal_00} we do not expect the networks we consider to appear in primary bifurcations from a trivial state - the symmetries we require for robustness do not act transitively on the set of cells.

We do not expect it will be possible to remove the constraint of no 1-cycles that is required by both constructions, except by lifting any graph with 1-cycles to a larger graph that has no 1-cycles. Because of the constraints of transversality, it is not possible for a dissipative system to have a robust homoclinic cycle to a hyperbolic equilibrium, even if the perturbations are constrained to leave certain subspaces invariant. This is because, within the smallest invariant subspace that contains the connection, there will be a transverse intersection that will be broken by small perturbations.

\subsection{Residence times near vertices}

For a given typical trajectory we define $I_k(j)$ to be the probability that the trajectory is near $\xi_j$, that is:
\[
I_k(j)=\{l\in \{1,\ldots,k\} : i_l=j\} ~\mbox{ and define }~N_k(j)=\#\{I_k(j)\}
\]
to be the cardinality of $I_k(j)$, i.e. the number of times a trajectory visits the equilibria $\xi_j$. This can be used to define the {\em mean residence time} of the trajectory close to the vertex $\xi_j$:
$$
\tau_{j}=\lim_{k\rightarrow\infty} \frac{1}{N_k(j)}\sum_{l\in I_k(j)} T_l.
$$
In this paper we do not examine properties of $\tau_j$ except to remark that this is a very interesting and nontrivial question \cite{AshOroWorTow07}. It was shown in~\cite{StoArm99} that the mean passage time of a trajectory past an equilibrium (that is, the length of time during which the trajectory remains in a neighbourhood of an equilibrium) for a heteroclinic cycle in the limit of low noise $\zeta\rightarrow 0$ is
\[
\tau_j=\frac{1}{\lambda_j}\log\left(\frac{h}{\zeta}\right)+O(1)
\]
where $\lambda_j$ is the largest expanding eigenvalue for equilibrium $\xi_j$ and $h$ is the neighbourhood size. For a  heteroclinic network in the presence of noise, the passage time will be similar, though if there are several expanding eigenvalues, it is possible that the transition time may depend on the exit route, in this way the mean residence times may in fact encode the perturbations of the system to inputs with non-zero mean \cite{wordsworth_ashwin_08}.

\subsection{Relevance to computational systems}

The constructions presented here should give flexible tools for designing coupled cell systems that realise finite-state computational systems. They suggest new ways to adapt coupled cell systems systems to ``learn'' networks, by modifying parameters \cite{AshOroWorTow07}. An interesting question is whether either of the constructions can adapted to explain neural computations that proceed on a dynamical basis. Simple network models of coupled neurons \cite{ashwin_karabacak_nowotny_2011} can give rise to various structures of heteroclinic network in phase space and we believe the constructions here can be used to extend this to more realistic decision making systems.

In applications, it may also be desirable to have control over the statistical properties of how a trajectory moves around the heteroclinic network. We may also want to control the occurrence of memory effects in the network where transition probabilities depend on the recent history of the route taken around the network; the methods in Section~\ref{sec:stats} offers a route for doing this.

There are many further questions that one could ask about the resulting designed networks - these include, for example: Can one design a network that has not only the given network structure, but also a specified set of average residence times and/or transition probabilities (with or without ``memory'')? This is likely to be an interesting and challenging problem where inclusion of anisotropic noise may be important \cite{AshPod10}. Other questions concern the limits on memory effects within such systems and behaviour of the system at larger noise levels.

Finally, we should comment that the realisation methods here are not very efficient in terms of the dimension of the system - the phase space dimension scales linearly with the number of vertices (resp. edges) for the simplex (resp. cylinder) realisations. This could be a barrier to using these results as a paradigm for neural computation where the encoding may be very dense. By contrast, the number of vertices for ``odd graph'' networks is very efficient - using only $n=2k+1$ globally coupled phase oscillators we can find heteroclinic networks with $n!/(k!(k+1)!)$ vertices \cite{AshOroBor_2010}. However, as previously highlighted, the latter networks are not easily adaptable to computation because of their complex topology and large number of symmetries. Hence it is interesting challenge to find a way to robustly realise a noisy heteroclinic attractor in a ``minimal'' dimension network.

\subsection*{Acknowledgments}

This work was started during discussions at the Mathematical Biosciences Institute (MBI) in Ohio in 2011 and continued during a visit of CMP to Exeter. We thank the MBI, the Royal Society and the University of Auckland for partial support for this research and thank Marc Timme, Fabio Neves, Chris Bick, Mike Field, Vivien Kirk and Ilze Ziedins for very interesting conversations in relation to this work. Finally, we thank the referees for advice that helped us to improve the exposition in the paper.

\bibliographystyle{plain}

\end{document}